\documentstyle[12pt]{article}
\input epsf
\newcommand{\nc}{\newcommand}
\nc{\postscript}[2] 
{\setlength{\epsfxsize}{#2\hsize}\centerline{\epsfbox{#1}}}
\nc{\bg}{B. Grz\c{a}dkowski}
\nc{\non}{\nonumber}
\nc{\barx}{\bar{x}}\nc{\pbarn}{\;\hbox {pb}}\nc{\fbarn}{\;\hbox {fb}}
\nc{\hc}{\hbox {h.c.}} 
\nc{\re}{\hbox {Re}} 
\nc{\im}{\hbox {Im}}
\nc{\mev}{\hbox {MeV}} \nc{\gev}{\;\hbox {GeV}}
\def\gesim{\lower0.5ex\hbox{$\:\buildrel >\over\sim\:$}} 
\def\lesim{\lower0.5ex\hbox{$\:\buildrel <\over\sim\:$}} 
\nc{\prd}[3]{{\it Phys.\ Rev.}\ {{\bf D{#1}} (#2), #3}}
\nc{\prl}[3]{{\it Phys.\ Rev.\ Lett.}\ {{\bf {#1}} (#2), #3}}
\nc{\plb}[3]{{\it Phys.\ Lett.}\ {{\bf B{#1}} (#2), #3}}
\nc{\npb}[3]{{\it Nucl.\ Phys.}\ {{\bf B{#1}} (#2), #3}}
\nc{\ptp}[3]{{\it Prog.\ Theor.\ Phys.}\ {{\bf {#1}} (#2), #3}}
\nc{\zfp}[3]{{\it Z.\ Phys.}\ {{\bf C{#1}} (#2), #3}}
\nc{\mpla}[3]{{\it Mod.\ Phys.\ Lett.}\ {{\bf A{#1}} (#2), #3}}
\nc{\rmp}[3]{{\it Rev.\ Mod.\ Phys.}\ {{\bf {#1}} (#2), #3}}
\nc{\ijmpa}[3]{{\it Int.\ J.\ of\ Mod.\ Phys.}\
               {{\bf A{#1}} (#2), #3}}
\nc{\ttbar}{t\bar{t}}         \nc{\bbbar}{b\bar{b}}
\nc{\tanb}{\tan \beta}        \nc{\twbdec}{t\to W^+ b}
\nc{\tbwbdec}{\bar{t}\to W^- \bar{b}}
\nc{\epem}{e^+e^-}            \nc{\eett}{\epem \to \ttbar}
\nc{\sigeett}{\sigma_{e\bar{e}\to\ttbar}}
\nc{\wpwm}{W^+W^-}            \nc{\tbar}{\bar{t}}
\nc{\bbar}{\bar{b}}           \nc{\wpp}{W^+}
\nc{\mt}{m_t}    \nc{\mts}{m_t^2}   \nc{\mw}{m_W}    \nc{\mws}{m_W^2}
\nc{\mz}{m_Z}    \nc{\mzs}{m_Z^2}
\nc{\ttbardec}{\ttbar \to W^+W^-\bbbar}
\nc{\wwbb}{W^+W^-\bbbar}      \nc{\sm}{SM}
\nc{\cw}{\cos\theta_W}        \nc{\sw}{\sin\theta_W}
\nc{\sws}{\sin^2\theta_W}     \nc{\sig}{\sigma_{tot}}
\nc{\lp}{{\ell}^+}              \nc{\lm}{{\ell}^-}
\nc{\epsl}{\epsilon_L}        \nc{\cp}{C\!P}

\nc{\splus}{s_+}       \nc{\smin}{s_-}        \nc{\eps}{\epsilon}
\nc{\psp}{Ps_+}        \nc{\psm}{Ps_-}        \nc{\lsp}{ls_+}
\nc{\lsm}{ls_-}        \nc{\sss}{s_+s_-}      \nc{\m}{m_t}
\nc{\mq}{m_t^2}        \nc{\mr}{\frac{1}{\m}} \nc{\av}{A_{\gamma}}
\nc{\bv}{B_{\gamma}}   \nc{\az}{A_Z}          \nc{\bz}{B_Z}
\nc{\avs}{A_{\gamma}^2}\nc{\azs}{A_Z^2}       \nc{\bzs}{B_Z^2}
\nc{\dav}{\delta \! A_{\gamma}}   \nc{\dbv}{\delta \! B_{\gamma}}
\nc{\dcv}{\delta C_{\gamma}}      \nc{\ddv}{\delta \! D_{\gamma}}
\nc{\daz}{\delta \! A_Z}          \nc{\dbz}{\delta \! B_Z}
\nc{\dcz}{\delta C_Z}             \nc{\ddz}{\delta \! D_Z}
\nc{\dev}{\delta \! E_{\gamma}}   \nc{\dez}{\delta \! E_Z}
\nc{\dfv}{\delta \! F_{\gamma}}   \nc{\dfz}{\delta \! F_Z}
\nc{\rdav}{{\rm Re}(\delta \! A_{\gamma}) \:}
\nc{\rdbv}{{\rm Re}(\delta \! B_{\gamma}) \:}
\nc{\rdcv}{{\rm Re}(\delta C_{\gamma}) \:}
\nc{\rddv}{{\rm Re}(\delta \! D_{\gamma}) \:}
\nc{\rdaz}{{\rm Re}(\delta \! A_Z) \:}
\nc{\rdbz}{{\rm Re}(\delta \! B_Z) \:}
\nc{\rdcz}{{\rm Re}(\delta C_Z) \:}
\nc{\rddz}{{\rm Re}(\delta \! D_Z) \:}
\nc{\idav}{{\rm Im}(\delta \! A_{\gamma}) \:}
\nc{\idbv}{{\rm Im}(\delta \! B_{\gamma}) \:}
\nc{\idcv}{{\rm Im}(\delta C_{\gamma}) \:}
\nc{\iddv}{{\rm Im}(\delta \! D_{\gamma}) \:}
\nc{\idaz}{{\rm Im}(\delta \! A_Z) \:}
\nc{\idbz}{{\rm Im}(\delta \! B_Z) \:}
\nc{\idcz}{{\rm Im}(\delta C_Z) \:}
\nc{\iddz}{{\rm Im}(\delta \! D_Z) \:}
\nc{\cz}{(1+v_e^2)d\:\!'^2}         \nc{\ci}{v_ed\:\!'}
\nc{\ccz}{v_ed\:\!'^2}              \nc{\cci}{d\:\!'}
\nc{\lspace}{\;\;\;\;\;\;\;\;\;\;}  \nc{\llspace}{\lspace \lspace}

\nc{\beq}{\begin{equation}}   \nc{\eeq}{\end{equation}}
\nc{\bea}{\begin{eqnarray}}   \nc{\eea}{\end{eqnarray}}
\nc{\baa}{\begin{array}}      \nc{\eaa}{\end{array}}
\nc{\bit}{\begin{itemize}}    \nc{\eit}{\end{itemize}}
\nc{\ben}{\begin{enumerate}}  \nc{\een}{\end{enumerate}}
\nc{\bce}{\begin{center}}     \nc{\ece}{\end{center}}
\begin{document}
\pagestyle{empty} \setlength{\footskip}{2.0cm}
\setlength{\oddsidemargin}{0.5cm} \setlength{\evensidemargin}{0.5cm}
\renewcommand{\thepage}{-- \arabic{page} --}
\def\mib#1{\mbox{\boldmath $#1$}}
\def\bra#1{\langle #1 |}      \def\ket#1{|#1\rangle}
\def\vev#1{\langle #1\rangle} \def\dps{\displaystyle}
   \def\thebibliography#1{\centerline{REFERENCES}
     \list{[\arabic{enumi}]}{\settowidth\labelwidth{[#1]}\leftmargin
     \labelwidth\advance\leftmargin\labelsep\usecounter{enumi}}
     \def\newblock{\hskip .11em plus .33em minus -.07em}\sloppy
     \clubpenalty4000\widowpenalty4000\sfcode`\.=1000\relax}\let
     \endthebibliography=\endlist
   \def\sec#1{\addtocounter{section}{1}\section*{\hspace*{-0.72cm}
     \normalsize\bf\arabic{section}.$\;$#1}\vspace*{-0.3cm}}
\vspace*{-1.6cm}\noindent
\hspace*{11.cm}IFT-06-98\\
\hspace*{11.cm}TOKUSHIMA 98-02\\
\hspace*{11.cm}(hep-ph/9805318)\\

\vspace*{.5cm}

\begin{center}
{\large\bf Probing Top-Quark Couplings at Polarized NLC}
\end{center}

\vspace*{1.5cm}
\begin{center}
\renewcommand{\thefootnote}{\alph{footnote})}
{\sc Bohdan GRZ\c{A}DKOWSKI$^{\:1),\:}$}\footnote{E-mail address:
\tt bohdan.grzadkowski@fuw.edu.pl}\ and\ 
{\sc Zenr\=o HIOKI$^{\:2),\:}$}\footnote{E-mail address:
\tt hioki@ias.tokushima-u.ac.jp}
\end{center}

\vspace*{1.2cm}
\centerline{\sl $1)$ Institute for Theoretical Physics,\ Warsaw 
University}
\centerline{\sl Ho\.za 69, PL-00-681 Warsaw, POLAND}

\vskip 0.3cm
\centerline{\sl $2)$ Institute of Theoretical Physics,\ 
University of Tokushima}
\centerline{\sl Tokushima 770-8502, JAPAN}

\vspace*{2.7cm}
\centerline{ABSTRACT}

\vspace*{0.4cm}
\baselineskip=20pt plus 0.1pt minus 0.1pt
The energy spectrum of the lepton(s) in $e^+e^-\!\to t\bar{t}\to
{\ell}^\pm \cdots /{\ell}^+{\ell}^- \cdots$ at next linear colliders
(NLC) is studied for arbitrary longitudinal beam polarizations as a
test of possible new physics in top-quark couplings. The most general
non-standard form factors are assumed for $\gamma \ttbar$, $Z\ttbar$
and $Wtb$ vertices to analyze new-physics effects in a
model-independent way. Expected precision in determining these form
factors is estimated applying the optimal-observable procedure to
the spectrum.

\vfill
\newpage
\renewcommand{\thefootnote}{\sharp\arabic{footnote}}
\pagestyle{plain} \setcounter{footnote}{0}
\baselineskip=21.0pt plus 0.2pt minus 0.1pt

\sec{Introduction}

The discovery of the top quark has completed the fermion spectrum
required by the electroweak standard model (SM). It is still an open
question, however, if the top-quark interactions obey the SM scheme
or there exists any new-physics contribution. The top quark decays
immediately after being produced \cite{Bigi} since its huge mass
$m_t^{exp}=175.6\pm 5.5$ GeV \cite{data97} leads to a decay width
${\mit\Gamma}_t$ much larger than ${\mit\Lambda}_{\rm QCD}$.
Therefore the decay process is not influenced by any fragmentation
effects and the decay products carry lots of information on the
top-quark properties. 

The energy distribution of the final lepton(s) in $e^+e^-\!\to t
\bar{t}\to {\ell}^\pm \cdots /{\ell}^+{\ell}^- \cdots$ turns out to
be a useful tool to analyze the top-quark couplings \cite{SP}. Indeed
it has been frequently studied in the literature over the past
several years \cite{CKP}--\cite{Rest} in order to find observables
sensitive to $C\!P$ violation. To illustrate this point, it will be
instructive to see how the spectrum is affected by non-conservation
of $C\!P$ in the production process: \\
Since $t\bar{t}$ are produced mainly through $\gamma/Z$ exchange,
their helicities would be only $(+-)$ or $(-+)$ if $m_t$ were much
smaller than $\sqrt{s}$. Fortunately, however, this is not the case
and we can expect copious $(++)$ and $(--)$ productions as well even
at $\sqrt{s}=500$ GeV.\footnote{A rough estimate within the SM gives
    $N(-+):N(+-):N(--):N(++)$ is $5:3.5:1:1$, where $N(\cdots)$
    denotes the number of $\ttbar$ pairs with the indicated
    helicities}\ 
These states transform into each other under $C\!P$ operation as
$\hat{C}\!\hat{P}\ket{\mp\mp}=\ket{\pm\pm}$, which means that the
difference $N(--)-N(++)$ could be a useful measure of $C\!P$
violation \cite{SP}--\cite{BG}. This important information cannot
be drawn directly since the top decays too rapidly as mentioned, but
is transferred to the final-lepton-energy distributions as follows:
\begin{itemize}
\item[(1)] The heavy top requires a large fraction $(\sim 70\,\%)$ of
$W$ bosons are longitudinally polarized in $t\to bW$ since $\bar{b}
\gamma_{\mu}\gamma_5 t\cdot\varepsilon^{\mu}\sim m_t\bar{b}\gamma_5
t$ when $\varepsilon^{\mu}=\varepsilon_L^{\mu}\sim k^{\mu}$
($\varepsilon$ and $k$ are the polarization and the four-momentum of
$W$, respectively).
\vspace*{-0.3cm}
\item[(2)] The produced $b$ ($\bar{b}$) is left-handed (right-handed)
in the SM since $m_b/\sqrt{s}\ll$1.
\vspace*{-0.3cm}
\item[(3)] Because of (1) and (2), $W^+$'s three-momentum prefers to
be parallel (anti-parallel) to that of $t(+)$($t(-)$), where
$t(\cdots)$ expresses a top with the indicated helicity. Consequently
${\ell}^+$ in the $t(+)$ decay becomes more energetic than in the
$t(-)$ decay, while it is just opposite for the $\bar{t}$ decay,
i.e., $\bar{t}(-)$ produces more energetic ${\ell}^-$ than $\bar{t}
(+)$ does.
\vspace*{-0.3cm}
\item[(4)] Therefore, we expect larger number of energetic ${\ell}^+$
(${\ell}^-$) for $N(--)<N(++)$ (for $N(--)>N(++)$).
\end{itemize}

In realistic analyses, one should take into account that other source
of non-SM effects may also exist. However, most of the
above-mentioned articles focused on $C\!P$-violating effects in
$\gamma/Z\ttbar$ vertices (production) only and did not assume the
most general form for the interactions of $\gamma\ttbar$, $Z\ttbar$
and $Wtb$. Therefore, in our previous paper \cite{BGH}, we have
performed a comprehensive analysis taking into account
$C\!P$-violating {\it and} $\cp$-conserving non-standard top-quark
couplings contributing both to the production {\it and} decay
process.
 
In this paper, extending that work for arbitrary longitudinal 
$e^\pm$ polarizations, we present a systematic way to determine the
non-SM parameters describing the general $\gamma/Z\ttbar$ and $Wtb$
couplings. In our another recent paper \cite{GHS} we have discussed
how the same process receives non-SM contributions from effective
four-Fermi interactions. Therefore, with the present work we will
complete a full analysis of anomalous effects in top-quark
interactions for polarized $e^+e^-$ beams in a model-independent way,
where beyond-the-SM physics is parameterized by the $SU(3)\times
SU(2)\times U(1)$ symmetric effective Lagrangian
\cite{effective_lag}.

This paper is organized as follows. First in sec.2 we describe the
basic framework of our analysis, and give the normalized single- and
double-lepton-energy distributions. Then, in sec.3, we estimate to
what precision all the non-standard parameters can be measured using
the optimal-observable method \cite{optimalization}. Adopting two
sets of non-SM-parameter values we show in detail how effective the
use of polarized beams could be for achieving better precision.
Finally, we summarize our results in sec.4. In the appendix we
collect several functions used in the main text for completeness,
though they could also be found in our previous papers
\cite{GH_npb,BGH,GHS}.

\sec{The lepton-energy distributions}

In this section we briefly present our formalism, and then derive
thereby the single- and double-lepton-energy distributions.

We will treat all the fermions except the top quark as massless and
adopt the technique developed in \cite{technique}. This is a useful
method to calculate distributions of final particles appearing in a
production process of on-shell particles and their subsequent decays.
This technique is applicable when the narrow-width approximation
$$
\left|\,{1\over{p^2-m^2+im{\mit\Gamma}}}\,\right|^2
\simeq{\pi\over{m{\mit\Gamma}}}\delta(p^2 -m^2)
$$
can be adopted for the decaying intermediate particles. In fact, this
is very well satisfied for both $t$ and $W$ since ${\mit\Gamma}_t
\simeq$ 175$(\mt/\mw)^3\;\mev \ll\mt$ and ${\mit\Gamma}_W\simeq 2$
GeV $\ll M_W$.

Adopting this method, one can derive the following formulas for the
inclusive distributions of the single-lepton ${\ell}^+$ and
double-lepton ${\ell}^+{\ell}^-$ in the reaction $\eett$\ \cite{AS}:
\begin{eqnarray}
&&\frac{d^3\sigma}{d^3\mib{p}_{\ell}/(2p_{\ell}^0)}
(\epem \to {\ell}^+ + \cdots)
\non \\
&&\ \ \ \ \ \ \ \ \ \ \ \ \ \ \ \ \ 
=\frac4{{\mit\Gamma}_t}\int d{\mit\Omega}_t
\frac{d\sigma}{d{\mit\Omega}_t}(n,0)
\frac{d^3{\mit\Gamma}_{\ell}}{d^3\mib{p}_{\ell}/(2p_{\ell}^0)}
(t\to b{\ell}^+\nu),
\label{master1}
\end{eqnarray}
\begin{eqnarray}
&&\frac{d^6\sigma}{d^3\mib{p}_{\ell}/(2p_{\ell}^0)
                   d^3\mib{p}'_{\ell}/(2{p_{\ell}^0}')}
(\epem \to {\ell}^+ {\ell}^- + \cdots)
\non \\
&&\ \ \ \
=\frac4{{\mit\Gamma}_t^2}
\int d{\mit\Omega}_t \frac{d\sigma}{d{\mit\Omega}_t}(n,m)
\frac{d^3{\mit\Gamma}_{\ell}}{d^3\mib{p}_{\ell}/(2p_{\ell}^0)}
(t\to b{\ell}^+\nu)
\frac{d^3{\mit\Gamma}_{\ell}}{d^3\mib{p}'_{\ell}/(2{p_{\ell}^0}')}
(\bar{t}\to \bar{b}{\ell}^-\bar{\nu}),\ \ 
\label{master2}
\end{eqnarray}
where ${\mit\Gamma}_{\ell}$ and ${\mit\Gamma}_t$ are the leptonic and
total widths of {\it unpolarized} top respectively, and
$d\sigma(n,m)/d{\mit\Omega}_t$ is obtained from the angular
distribution of $\ttbar$ with spins $s_+$ and $s_-$ in $\eett$,
$d\sigma(s_+,s_-)/d{\mit\Omega}_t$, by the following replacement:
\begin{eqnarray}
&&s^\mu_+ \to n^\mu=
+\left(g^{\mu \nu}-\frac{p_t^\mu p_t^\nu}{\mts}
\right)\frac{\mt}{p_t p_{\ell}}p_{{\ell}\,\nu} \nonumber \\
&&s^\mu_- \to m^\mu=
-\left(g^{\mu \nu}-\frac{p_{\bar{t}}^\mu p_{\bar{t}}^\nu}{\mts}
\right)\frac{\mt}{p_{\bar{t}} p'_{\ell}}p'_{{\ell}\,\nu}.
\label{replacement}
\end{eqnarray}
(Exchanging the roles of $s_+$ and $s_-$ and reversing the sign of
$n^\mu$, we get the single distribution of ${\ell}^-$.)

In order to obtain the lepton spectra according to the above formulas
we shall first calculate the $t\bar{t}$-production cross section and
their decay rates.

\noindent
$\mib{t}\bar{\mib{t}}$ {\bf production}

Let us start with the $t\bar{t}$ production. We can represent the
most general $\ttbar$ couplings to the photon and $Z$ boson as
\begin{equation}
{\mit\Gamma}_{vt\bar{t}}^{\mu}=
\frac{g}{2}\,\bar{u}(p_t)\,\Bigl[\,\gamma^\mu \{A_v+\delta\!A_v
-(B_v+\delta\!B_v) \gamma_5 \}
+\frac{(p_t-p_{\bar{t}})^\mu}{2m_t}(\delta C_v-\delta\!D_v\gamma_5)
\,\Bigr]\,v(p_{\bar{t}}),\ \label{ff}
\end{equation}
where $g$ denotes the $SU(2)$ gauge coupling constant, $v=\gamma,Z$,
and
\[
\av=\frac43\sw,\ \ \bv=0,\ \ 
\az=\frac1{2\cw}\Bigl(1-\frac83\sin^2\theta_W\Bigr),\ \ 
\bz=\frac1{2\cw}.
\]
Among the above form factors, $\delta\!A_{\gamma,Z}$, $\delta\!
B_{\gamma,Z}$, $\delta C_{\gamma,Z}$ and $\delta\!D_{\gamma,Z}$ are
parameterizing $C\!P$-conserving and $C\!P$-violating non-standard
interactions, respectively. Note that we dropped two other
independent terms proportional to $(p_t +p_{\bar{t}})^\mu$ since
their effects vanish in the limit of zero electron mass.

On the other hand, interactions of initial $\epem$ have been assumed
untouched by non-standard interactions since their structures are
well described within the SM:
\begin{itemize}
\item $\gamma e^+ e^-$ vertex
\beq
{\mit\Gamma}^{\mu}_{\gamma e^+e^-} =
-e\:\bar{v}(p_{e^+})\,\gamma^{\mu}\,u(p_{e^-})\,,
\eeq
\item $Z e^+e^-$ vertex
\beq
{\mit\Gamma}^{\mu}_{Z e^+e^-} = \frac{g}{4 \cw}\,
\bar{v}(p_{e^+})\,\gamma^{\mu}(v_e+\gamma_5)\,u(p_{e^-})\,,
\eeq
\end{itemize}
where $v_e\equiv -1+4\sin^2\theta_W$.

The angular distribution of polarized $t\bar{t}$ pair in presence
of the above non-standard interactions is obtained after a tedious
but straightforward calculation. The result is however a bit too
lengthy, so we give the explicit form in the appendix and here
instead we describe its structure rather qualitatively: \\
First, the invariant amplitude can be expressed as
\begin{equation}
{\cal M}=\sum_{i,I}C_{iI}\:j_\mu^i J^{I\mu}
\end{equation}
where
\begin{eqnarray*}
&&j_\mu^i \equiv \bar{v}(p_{e^+}){\mit\Gamma}^i_\mu u(p_{e^-})\ \ \
(i=V,A) \\
&&J_\mu^I \equiv \bar{u}(p_t){\mit\Gamma}^I_\mu v(p_{\bar{t}})\ \ \
(I=V,A,S,P) 
\end{eqnarray*}
and
$$
{\mit\Gamma}_\mu^{V,A,S,P}\equiv\gamma_\mu,\ \gamma_\mu\gamma_5,\
q_\mu,\ q_\mu\gamma_5\ \ \ (q\equiv p_t -p_{\bar{t}}).
$$
Therefore $|{\cal M}|^2$ consists of a number of terms whose
coefficients are $C_{iI}^*C_{i'I'}$. In the explicit formula in the
appendix, we express $C_{iI}^*C_{iI'}$ $(I,I'=V,A)$, $C_{iI}^*
C_{i'I'}$ $(i\neq i'\ {\rm and}\ I,I'=V,A)$, $C_{iI}^*C_{i'P}$
$(i,i',I=V,A)$ and $C_{iI}^*C_{i'S}$ $(i,i',I=V,A)$ as $D$, $E$, $F$
and $G$ respectively, and moreover we attach subscripts $V$, $A$ and
$V\!\!A$ to $D$ and $E$ according to $[I=I'=V]$, $[I=I'=A]$ and
$[I=V,I'=A]$, while $F$ and $G$ are classified by $i=1\sim 4$
according to their $V/A$ structure.\footnote{More explicit formulas
    will appear in a separate paper \cite{HO}.}

It is worth to notice that: \\
$\bullet$ In the SM-limit only $D_{V,A,V\!\!A}$ and $E_{V,A,V\!\!A}$
remain and all $F_i=G_i=0$, \\
$\bullet$ Non-zero $F_i$'s are generated by the $C\!P$-violating form
factors $\delta\!D_{\gamma,Z}$, \\
$\bullet$ Contributions to $G_i$'s are created by the
$C\!P$-conserving form factors $\delta C_{\gamma,Z}$.

For the initial beam-polarization we follow the convention by Tsai
\cite{Tsai}:
\bea
&&P_{e^-}=+[N(e^-,+1)-N(e^-,-1)]/[N(e^-,+1)+N(e^-,-1)],\\
&&P_{e^+}=-[N(e^+,+1)-N(e^+,-1)]/[N(e^+,+1)+N(e^+,-1)],
\eea
where $N(e^{-(+)},h)$ is the number of $e^-(e^+)$ with helicity $h$
in each beam.\footnote{Note that $P_{e^+}$ is defined with the
    opposite overall sign in some other papers (see, e.g.,
    \cite{Blon}).}\ 
When the initial $e^-$ and $e^+$ get polarized, $j_\mu^V$ and
$j_\mu^A$ mix with each other since the spin (helicity) projection
operator for $u(p_{e^-})$ and $v(p_{e^+})$ in the massless limit is
$(1\pm\gamma_5)/2$. Then we obtain the cross section for
arbitrarily-polarized $e^+e^-$ beams by replacing $D_V$, $D_A$,
$D_{V\!\!A}$, $E_V$, $E_A$, $E_{V\!\!A}$, $F_i$ and $G_i$ $(i=1\sim
4)$ with $D_V^{(*)}$, $D_A^{(*)}$, $D_{V\!\!A}^{(*)}$, $E_V^{(*)}$,
$E_A^{(*)}$, $E_{V\!\!A}^{(*)}$, $F_i^{(*)}$ and $G_i^{(*)}$, where
\begin{eqnarray*}
&&D_{V,\,A,\,V\!\!A}^{(*)}=(1+P_{e^-}P_{e^+})D_{V,\,A,\,V\!\!A}
-(P_{e^-}+P_{e^+})E_{V,\,A,\,V\!\!A}, \\
&&E_{V,\,A,\,V\!\!A}^{(*)}=(1+P_{e^-}P_{e^+})E_{V,\,A,\,V\!\!A}
-(P_{e^-}+P_{e^+})D_{V,\,A,\,V\!\!A}, \\
&&F_{1,\,2,\,3,\,4}^{(*)}=(1+P_{e^-}P_{e^+})F_{1,\,2,\,3,\,4}
-(P_{e^-}+P_{e^+})F_{2,\,1,\,4,\,3}, \\
&&G_{1,\,2,\,3,\,4}^{(*)}=(1+P_{e^-}P_{e^+})G_{1,\,2,\,3,\,4}
-(P_{e^-}+P_{e^+})G_{2,\,1,\,4,\,3}.
\end{eqnarray*}

\noindent
{\bf $\mib{t}$ and $\bar{\mib{t}}$ decays}

We will adopt the following parameterization of the $Wtb$ vertex
suitable for the $\twbdec$ and $\tbwbdec$ decays:
\begin{eqnarray}
&&\!\!{\mit\Gamma}^{\mu}_{Wtb}=-{g\over\sqrt{2}}V_{tb}\:
\bar{u}(p_b)\biggl[\,\gamma^{\mu}(f_1^L P_L +f_1^R P_R)
-{{i\sigma^{\mu\nu}k_{\nu}}\over M_W}
(f_2^L P_L +f_2^R P_R)\,\biggr]u(p_t),\ \ \ \ \ \ \label{ffdef}\\
&&\!\!\bar{\mit\Gamma}^{\mu}_{Wtb}=-{g\over\sqrt{2}}V_{tb}^*\:
\bar{v}(p_{\bar{t}})
\biggl[\,\gamma^{\mu}(\bar{f}_1^L P_L +\bar{f}_1^R P_R)
-{{i\sigma^{\mu\nu}k_{\nu}}\over M_W}
(\bar{f}_2^L P_L +\bar{f}_2^R P_R)\,\biggr]v(p_{\bar{b}}),
\label{ffbdef}
\end{eqnarray}
where $P_{L/R}=(1\mp\gamma_5)/2$, $V_{tb}$ is the $(tb)$ element of
the Kobayashi-Maskawa matrix and $k$ is the momentum of $W$. Because
$W$ is on shell,\footnote{Remember that we use the narrow-width
    approximation also for the $W$ propagator.}\ 
the two additional form factors were not taken into account. It is
worth to mention that the above form factors satisfy the following
relations \cite{cprelation}:
\begin{equation}
f_1^{L,R}=\pm\bar{f}_1^{L,R},\lspace f_2^{L,R}=\pm\bar{f}_2^{R,L},
\label{cprel}
\end{equation}
where the upper (lower) signs are those for $C\!P$-conserving
(-violating) contributions.\footnote{Assuming $C\!P$-conserving
    Kobayashi-Maskawa matrix.}\ 

$Wl\nu$ couplings are treated within the SM as $\gamma/Z\ttbar$
couplings:
\bea
&&{\mit\Gamma}^\mu_{Wl\nu}=-\frac{g}{2\sqrt{2}}\,
  \bar{u}(p_\nu)\gamma^\mu(1-\gamma_5)v(p_{{\ell}^+}), \\
&&\bar{\mit\Gamma}^\mu_{Wl\nu}=-\frac{g}{2\sqrt{2}}\,
  \bar{u}(p_{{\ell}^-})\gamma^\mu(1-\gamma_5)v(p_{\bar{\nu}}).
\eea

Assuming that
$\stackrel{\scriptscriptstyle(-)}{f_1^L}-1$,
$\stackrel{\scriptscriptstyle(-)}{f_1^R}$,
$\stackrel{\scriptscriptstyle(-)}{f_2^L}$ and
$\stackrel{\scriptscriptstyle(-)}{f_2^R}$ are small and keeping only
their linear terms, we obtain for the differential spectrum the
following result:
\begin{equation}
\frac{1}{{\mit\Gamma}_t}
\frac{d^2{\mit\Gamma}_{\ell}}{dxd\omega}(t\to b{\ell}^+ \nu)
=\frac{1+\beta}{\beta}\;\frac{3 B_{\ell}}{W}
\omega\left[1+2{\rm Re}(f_2^R)\sqrt{r}\left(\frac{1}{1-\omega}-
\frac{3}{1+2r} \right)\right], \label{t-decay}
\end{equation}
where $x$ is the rescaled lepton-energy introduced in \cite{AS}
$$
x\equiv
\frac{2 E_{\ell}}{\mt}\left(\frac{1-\beta}{1+\beta}\right)^{1/2},
$$
with $E_{\ell}$ being the energy of ${\ell}$ in $\epem$ c.m. frame,
$\omega$ is defined as
$$
\omega\equiv(p_t -p_{\ell})^2/m_t^2,
$$
$B_{\ell}$ is the leptonic branching ratio of $t$ ($\simeq 0.22$ for
${\ell}=e,\mu$), and
$$
W\equiv(1-r)^2(1+2r),\ \ \ \ r\equiv(M_W/m_t)^2.
$$
An analogous formula holds for $\bar{t}\to\bar{b}{\ell}^-\bar{\nu}$
with $f_2^R$ replaced by $\bar{f}_2^L$.

\noindent
{\bf Lepton-energy distributions}

Now let us give the lepton-energy spectra in terms of $x$. Since we
are going to apply the method of optimal observables
\cite{optimalization} in order to isolate various non-standard
contributions, it is convenient to express the spectrum as a sum of
known independent functions multiplied by coefficients parameterizing
non-standard physics to be determined. In the following, we use as
input data $M_W=80.43$ GeV, $M_Z=91.1863$ GeV, $m_t=175.6$ GeV,
$\sin^2\theta_W=0.2315\:$\cite{data97}\ and $\sqrt{s}=$500 GeV.

\noindent
{\bf 1. Single distribution}

Adopting the formulas
eqs.(\ref{master1},\ref{distribution},\ref{t-decay}), keeping only
linear terms in non-standard form factors, and integrating over
${\mit\Omega}_t$ and the necessary top-quark-decay phase space, one
obtains the following normalized single-lepton-energy spectrum:
\beq
\frac1{B_{\ell} \sigma_{e \bar{e} \to t \bar{t}} }
{\frac{d\sigma}{dx}}^{\!\pm}=\sum_{i=1}^{3}c_i^\pm f_i(x),
\label{single}
\eeq
where $\sigeett\equiv\sigma_{tot}(e^+e^-\!\to\ttbar)$ and $\pm$
corresponds to ${\ell}^{\pm}$. The first term comes from the SM and
the coefficients are
$$
c_1^\pm =1,
$$
the second term originates from the anomalous $\gamma/Z\ttbar$
couplings (see eq.(\ref{ff})) contributing to the production process
$$
c_2^\pm = a_1\,\delta\!D_V^{(*)}
-a_2\,[\,\delta\!D_A^{(*)}-{\rm Re}(G_1^{(*)})\,]+a_3{\rm Re}
(\delta\!D_{V\!\!A}^{(*)})\mp \xi^{(*)},
$$
and the third term comes from the non-SM $Wtb$ couplings (see
eqs.(\ref{ffdef},\ref{ffbdef})) which influence the top-quark decay
distribution (see eq.(\ref{t-decay}))
$$
c_3^+={\rm Re}(f_2^R),\ \ c_3^-={\rm Re}(\bar{f}_2^L).
$$
Here, $\delta\!D_{V,A,V\!\!A}^{(*)}$ are the non-SM parts of
$D_{V,A,V\!\!A}^{(*)}$, $\xi^{(*)}$ is a $C\!P$-violating parameter
in the production process which is defined in a similar way as $\xi$
used in \cite{AS,GH_npb}:
$$
\xi^{(*)}\equiv 2\,{\rm Re}(F_1^{(*)})\,a_{V\!\!A}^{(*)},
$$
and $a_i$ are defined as
$$
a_1 \equiv -\,\eta^{(*)} (3-\beta^2)a_{V\!\!A}^{(*)},\ \ \
a_2 \equiv 2\,\eta^{(*)}\beta^2 a_{V\!\!A}^{(*)},\ \ \
a_3 \equiv 4\,a_{V\!\!A}^{(*)},
$$
with
$a_{V\!\!A}^{(*)}\equiv
1/[\,(3-\beta^2)D^{(0,\,*)}_V +2\beta^2 D^{(0,\,*)}_A\,]$ (the
superscript ``(0)" denotes the SM-part) and
$$
\eta^{(*)}\equiv 4\:a_{V\!\!A}^{(*)}D^{(0,\,*)}_{V\!\!A}
$$
($=0.2074$ in case of no beam polarization). On the other hand, the
functions $f_i(x)$ are
\beq
f_1(x)=f(x)+\eta^{(*)}\:g(x),\ \
f_2(x)=g(x),\ \
f_3(x)=\delta\!f(x)+\eta^{(*)}\:\delta g(x),
\label{fx}
\eeq
where
$f(x)$ and $g(x)$ are functions introduced in \cite{AS}, while
$\delta\!f(x)$ and $\delta g(x)$ are functions derived in our
previous work \cite{GH_npb}, which satisfy the following
normalization conditions:
$$
\int f(x)dx=1\ \ \ {\rm and}\ \ \ \int g(x)dx=\int \delta\!f(x)dx
=\int \delta g(x) dx=0.
$$
$f(x)$ and $g(x)$ describe the process with the standard top decays
while $\delta\!f(x)$ and $\delta g(x)$ come from the non-standard
contribution to the decay process. Here let us remind readers that
the $c_2^\pm$ term in (\ref{single}), which is proportional to
$g(x)$, originates in the spin dependent part of the lepton spectrum
and would vanish if, for instance, hadronization effects would dilute
the top-quark polarization. As explained in the introduction the
lepton-energy spectrum should depend on the polarization of the
parent top quark, that is the reason why all the non-standard effects
in the production process manifest themselves as modification of the
coefficient in front of $g(x)$ for the normalized spectrum. We
recapitulate these functions in the appendix.

It should be emphasized that the coefficients $c_i^\pm$ contain both
contributions from $\cp$-conserving and $\cp$-violating interactions,
therefore their determination does not provide a direct test of $\cp$
invariance. However, as was discussed in ref.~\cite{GH_npb} one can
easy combine measurements of the spectra for $\lp$ and $\lm$ in order
to construct purely $\cp$-violating observables like $d\sigma^+/dx-
d\sigma^-/dx$. It is also worth to notice here that even though
measurement of $c_i$ does not disentangle $\cp$-conserving and
$\cp$-violating interactions it allows for discrimination between
non-standard effects originating from the production and those from
the decay.

\begin{figure}[h]
\postscript{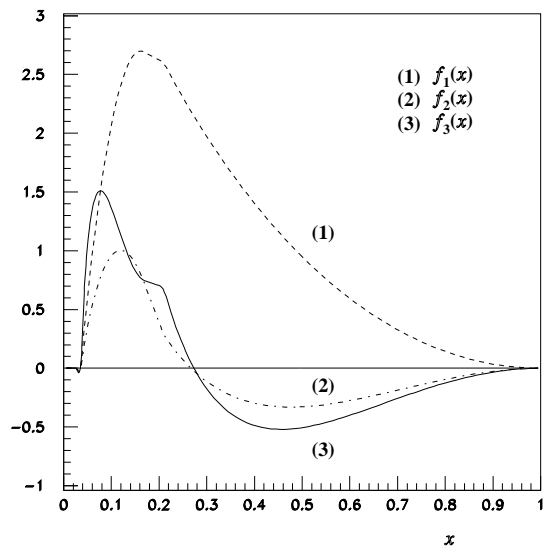}{0.6}
\vspace*{-0.5cm}
\caption{The functions $f_i(x)$ defined in eq.~(\protect\ref{fx})
for $P_{e^+}=P_{e^-}=0$.}
\label{plot2d}
\end{figure}

\begin{figure}
\postscript{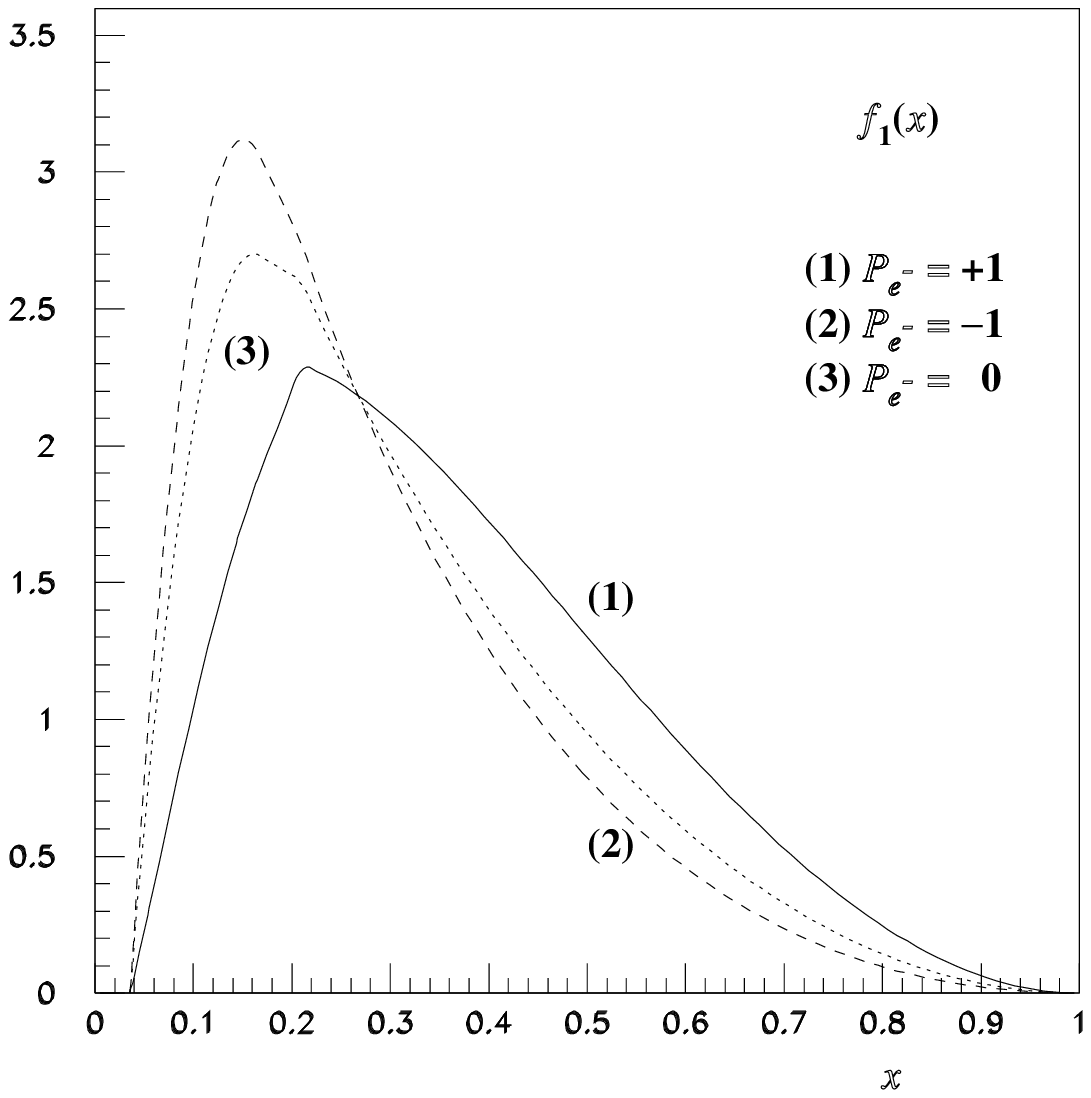}{0.6}
\vspace*{-0.5cm}
\caption{The function $f_1(x)$ for $P_{e^-}=+1$ vs $P_{e^+}=0/+1$
(solid line), for $P_{e^-}=-1$ vs $P_{e^+}=0/-1$ (dashed line) and
for no polarization (dotted line).}
\label{plot2d1p}
%
\postscript{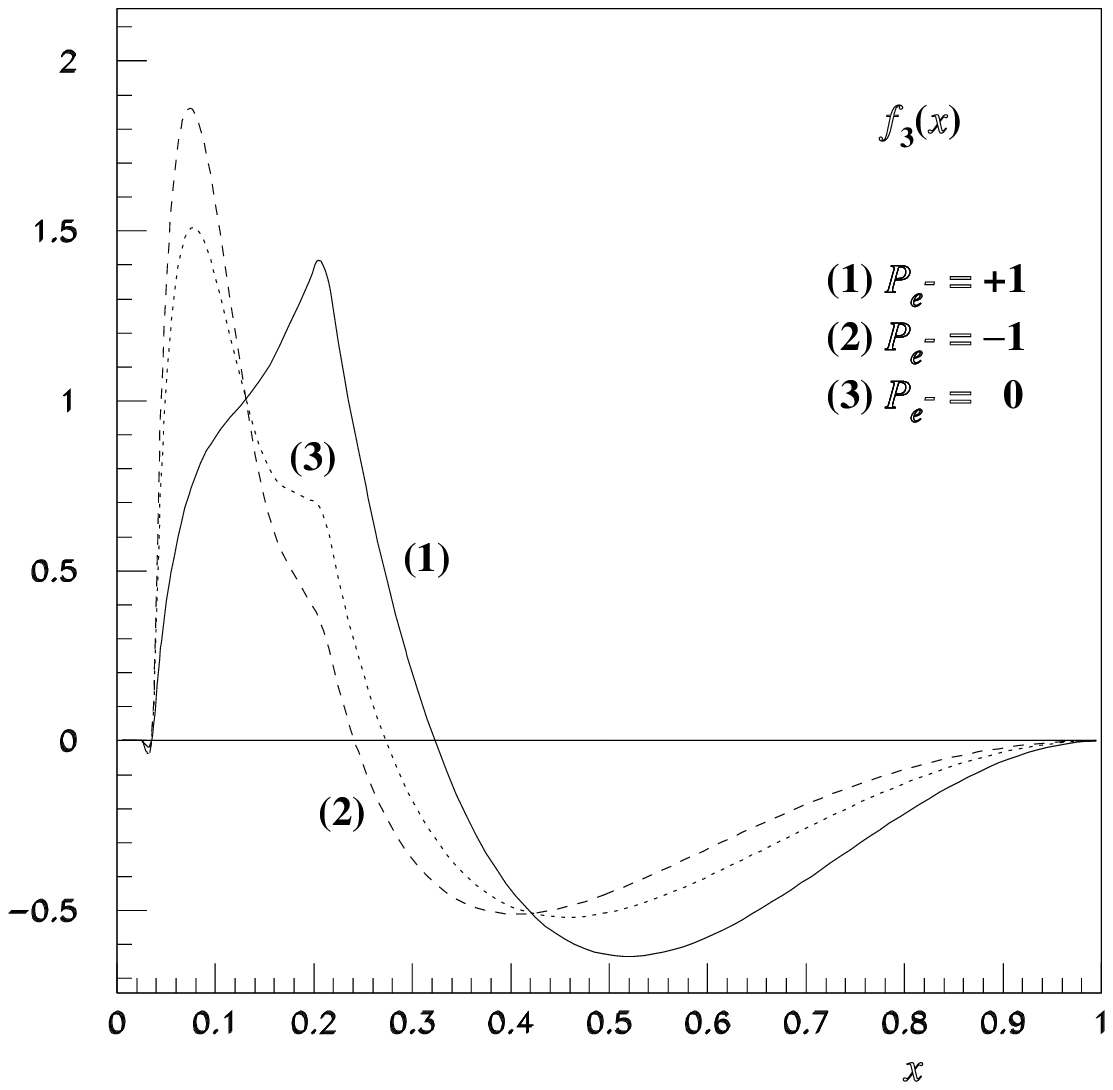}{0.6}
\vspace*{-0.5cm}
\caption{The function $f_3(x)$ for $P_{e^-}=+1$ vs $P_{e^+}=0/+1$
(solid line), for $P_{e^-}=-1$ vs $P_{e^+}=0/-1$ (dashed line) and
for no polarization (dotted line).}
\label{plot2d3p}
\end{figure}

The functions $f_i(x)$ are shown in fig.\ref{plot2d} for unpolarized
beams. Since $f_{1,3}(x)$ have $P_{e^\pm}$ dependence through
$\eta^{(*)}$, we also present them in figs.\ref{plot2d1p} and
\ref{plot2d3p} respectively for $P_{e^-}=+1$ vs $P_{e^+}=0/+1$ and
for $P_{e^-}=-1$ vs $P_{e^+}=0/-1$ as examples.\footnote{$P_{e^+}=
    0$ and $+1(-1)$ give the same $\eta^{(*)}$ and consequently the
    same $f_{1,3}(x)$ when $P_{e^-}=+1(-1)$.}
%
%
\clearpage

\noindent
{\bf 2. Double distribution}

Applying the same algorithm as for the single spectrum one finds for
the normalized double-lepton-energy spectrum the following formula:
\beq
\frac1{B_{\ell}^{2} \sigma_{e \bar{e} \to t \bar{t}} } 
\frac{d^2 \sigma}{dx d\bar{x}}\;=\;\sum_{i=1}^{6}c_i f_i(x,\bar{x}),
\label{double}
\eeq
where the first term comes from the SM
\[
c_1=1,
\]
the second and third terms are $C\!P$-violating non-SM contributions
of $\gamma/Z\ttbar$ and $Wtb$ couplings respectively,
\[
c_2=\xi^{(*)},\;\;c_3=\frac12{\rm Re}(f_2^R-\bar{f}_2^L),
\]
the fourth and fifth terms are both $C\!P$-conserving non-SM
$\gamma/Z\ttbar$ contributions
\[
c_4=a_1'\,\delta\!D_V^{(*)}+a_2'\,\delta\!D_A^{(*)}
+a_3'{\rm Re}(G_1^{(*)}),
\]
\[
c_5 = a_1\,\delta\!D_V^{(*)}
-a_2\,[\,\delta\!D_A^{(*)}-{\rm Re}(G_1^{(*)})\,]+a_3{\rm Re}
(\delta\!D_{V\!\!A}^{(*)}),
\]
while the last term is $C\!P$-conserving non-SM $Wtb$ contribution
\[
c_6=\frac12{\rm Re}(f_2^R+\bar{f}_2^L).
\]
The corresponding functions are
\begin{eqnarray}
f_1(x,\bar{x})&\!\!=&\!\!f(x)f(\bar{x})+\eta^{(*)}\:[\:f(x)g(\bar{x})
   +g(x)f(\bar{x})\:]+\eta'^{(*)}g(x)g(\bar{x}), \non \\ 
f_2(x,\bar{x})&\!\!=&\!\!f(x)g(\bar{x})-g(x)f(\bar{x}), \non \\ 
f_3(x,\bar{x})&\!\!=&\!\!\delta\!f(x)f(\bar{x})
   -f(x)\delta\!f(\bar{x})\non \\ 
&\!\!+&\!\!\eta^{(*)}\:[\:\delta\!f(x)g(\bar{x})
   -f(x)\delta g(\bar{x})
   +\delta g(x)f(\bar{x})-g(x)\delta\!f(\bar{x})\:] \non \\
&\!\!+&\!\!\eta'^{(*)}
   [\:\delta g(x)g(\bar{x})-g(x)\delta g(\bar{x})\:],
   \non \\ 
f_4(x,\bar{x})&\!\!=&\!\!g(x)g(\bar{x}),            \non \\
f_5(x,\bar{x})&\!\!=&\!\!f(x)g(\bar{x})+g(x)f(\bar{x}), \non \\
f_6(x,\bar{x})&\!\!=&\!\!\delta\!f(x)f(\bar{x})
   +f(x)\delta\!f(\bar{x}) \non \\ 
&\!\!+&\!\!\eta^{(*)}\:
   [\:\delta\!f(x)g(\bar{x})+f(x)\delta g(\bar{x}) 
   +\delta g(x)f(\bar{x})+g(x)\delta\!f(\bar{x})\:] \non \\
&\!\!+&\!\!\eta'^{(*)}
   [\:\delta g(x)g(\bar{x})+g(x)\delta g(\bar{x})\:], \label{fxxbar}
\end{eqnarray}
with $\eta'^{(*)}\equiv \beta^{-2}a_{V\!\!A}^{(*)}
[\,(1+\beta^2)D^{(0,\,*)}_V+2\beta^2 D^{(0,\,*)}_A\,](=1.2720$ for
$P_e =P_{\bar{e}}=0$) and $a_i'$ being defined as
\begin{eqnarray*}
&&a_1' \equiv
[\,\beta^{-2}(1+\beta^2)-(3-\beta^2)\eta'^{(*)}\,]\,a_{V\!\!A}^{(*)},
\\
&&a_2' \equiv 2(1-\beta^2 \eta'^{(*)})\,a_{V\!\!A}^{(*)},\ \ \
  a_3' \equiv 2(1+\beta^2\eta'^{(*)})\,a_{V\!\!A}^{(*)}.
\end{eqnarray*}
$f_{1,4,5,6}(x,\bar{x})$ and $f_{2,3}(x,\bar{x})$ are respectively
symmetric and antisymmetric in $(x,\bar{x})$, which are signals of
$C\!P$ conservation and $C\!P$ violation. Since $f_4$ and $f_5$ are
both from the $C\!P$-conserving parts of the production process, we
may recombine them, but we chose the above combination so that only
$f_5$ remains in computing the single distributions.

Here, as for the single spectrum, since for a given $c_i$ there is no
mixing between the production and decay processes, we will be able to
judge if the non-standard contributions originate from the production
or from the decay of top quarks. Furthermore, in contrast with the
single spectrum, the coefficients $c_i$ receive contributions either
from $\cp$-conserving ($i=1,4,5,6$) or $\cp$-violating ($i=2,3$)
interactions. Therefore determination of the coefficients provides a
direct test of $\cp$ invariance.

\begin{figure}
\vspace*{-0.5cm}
\postscript{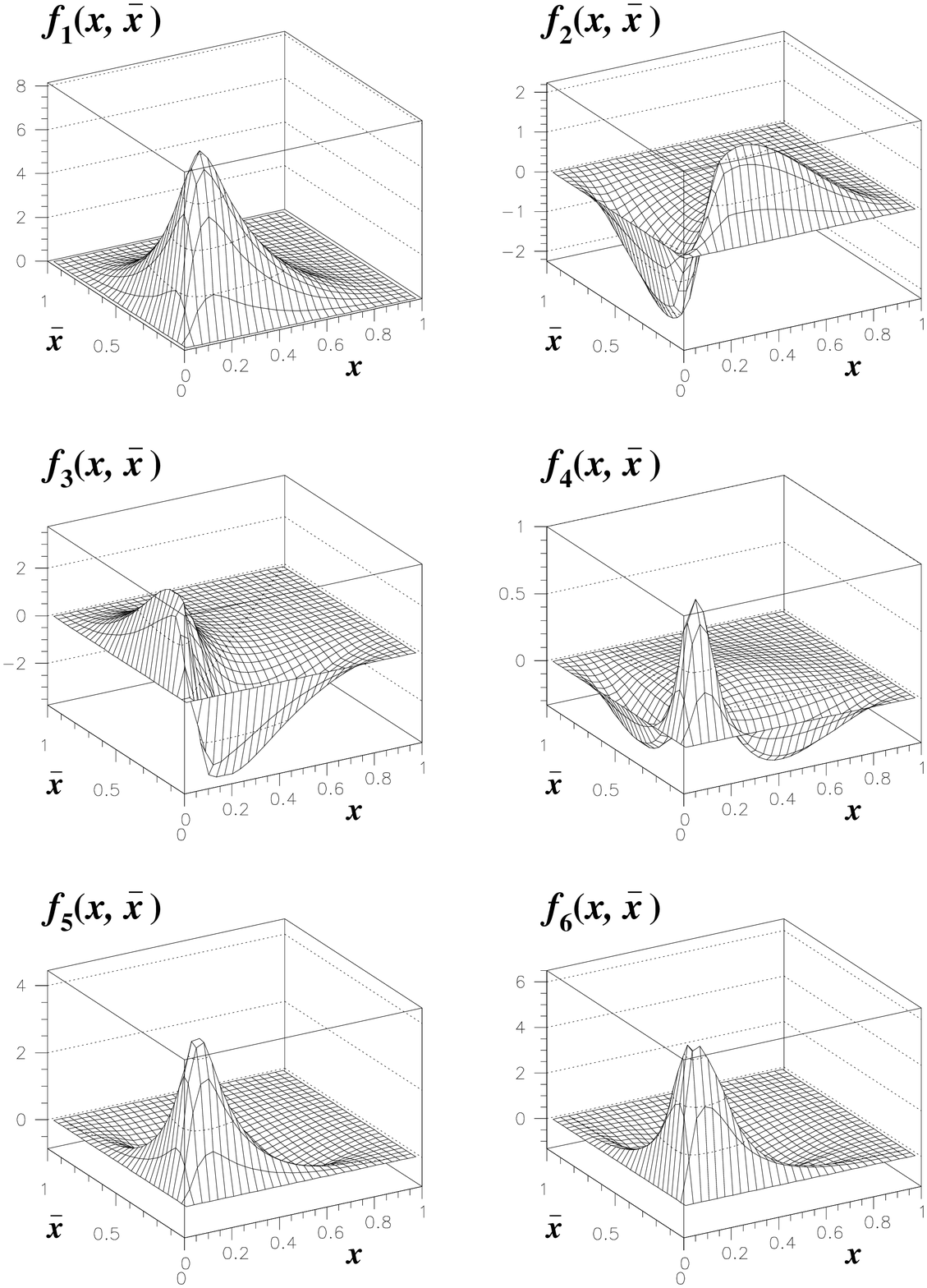}{1} 
\caption{The functions $f_i(x,\bar{x})$ defined in
eq.(\protect\ref{fxxbar}) for $P_{e^-}=P_{e^+}=0$.}
\label{plots}
\end{figure}

\begin{figure}
\postscript{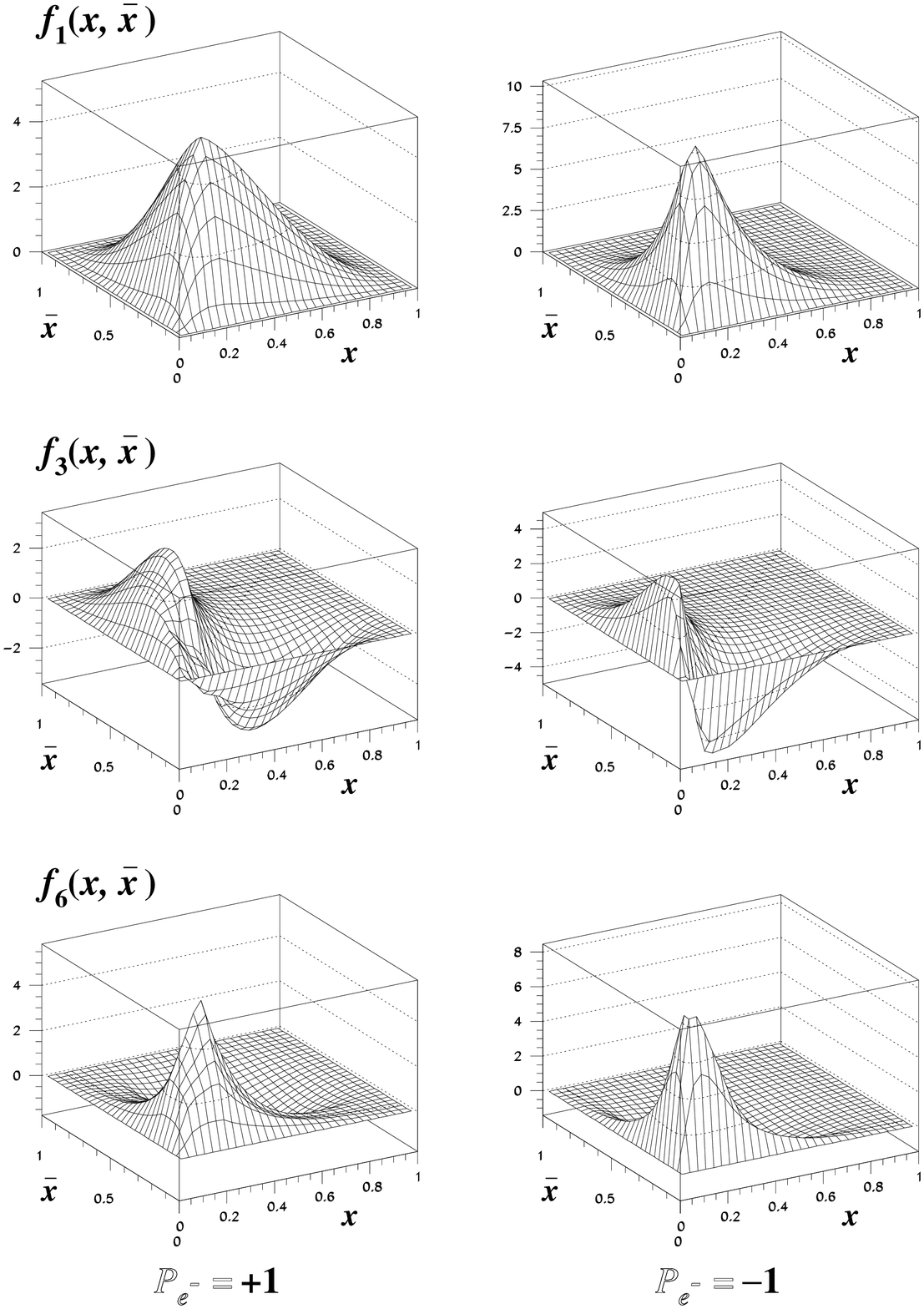}{1} 
\caption{The functions $f_{1,3,6}(x,\bar{x})$ for $P_{e^-}=+1$ vs
$P_{e^+}=0/+1$ (on the left side) and for $P_{e^-}=-1$ vs
$P_{e^+}=0/-1$ (on the right side).}
\label{plotsp}
\end{figure}

The functions $f_i(x,\bar{x})$ are plotted in fig.\ref{plots} for
unpolarized case. Since $f_{1,3,6}(x,\bar{x})$ depend on $P_{e^\pm}$
through $\eta^{(*)}$ and/or $\eta'^{(*)}$, we also show them in
fig.\ref{plotsp} for $P_{e^-}=+1$ vs $P_{e^+}=0/+1$ (on the left
side) and for $P_{e^-}=-1$ vs $P_{e^+}=0/-1$ (on the right side) as
examples. It can be observed from the figures that the shapes of the
functions $f_{1,3}(x)$ and $f_{1,3,6}(x,\bar{x})$ vary substantially
with the polarization of the initial beams. Therefore it is justified
to consider determination of the coefficients $c_i$ through
energy-spectrum measurements for various polarizations since one can
hope that carefully-adjusted beam-polarization may increase precision
of the analysis.\footnote{Getting higher statistics is also a reason
    for considering polarized beams.}

\newpage
\clearpage

\sec{The optimal observables}

We are now ready to perform a numerical analysis, but let us first
summarize the main points of the optimal-observable technique
\cite{optimalization}. Suppose we have a cross section
$$
\frac{d\sigma}{d\phi}(\equiv{\mit\Sigma}(\phi))=\sum_i c_i f_i(\phi)
$$
where $f_i(\phi)$ are known functions of the location in final-state
phase space $\phi$ and $c_i$'s are model-dependent coefficients. The
goal would be to determine  $c_i$'s. It can be done by using
appropriate weighting functions $w_i(\phi)$ such that $\int w_i(\phi)
{\mit\Sigma}(\phi)d\phi=c_i$. Generally, different choices for
$w_i(\phi)$ are possible, but there is a unique choice so that the
resultant statistical error is minimized. Such functions are given by
\begin{equation}
w_i(\phi)=\sum_j X_{ij}f_j(\phi)/{\mit\Sigma}(\phi)\,, \label{X_def}
\end{equation}
where $X_{ij}$ is the inverse matrix of $M_{ij}$ which is defined as
\begin{equation}
M_{ij}\equiv \int {f_i(\phi)f_j(\phi)\over{\mit\Sigma}(\phi)}d\phi\,.
\label{M_def}
\end{equation}
When we take these weighting functions, the statistical uncertainty
of $c_i$-\-deter\-mination through $d\sigma/d\phi$ measurement
becomes
\begin{equation}
{\mit\Delta}c_i=\sqrt{X_{ii}\,\sigma_T/N}\,, \label{delc_i}
\end{equation}
where $\sigma_T\equiv\int (d\sigma/d\phi) d\phi$ and $N$ is the total
number of events. It is clear from the definition of the matrix
$M_{ij}$, eq.(\ref{M_def}), that $M_{ij}$ has no inverse if the
functions $f_i(\phi)$ are linearly dependent, and then we cannot
perform any meaningful analysis. One can see it more intuitively as
follows: when $f_i(\phi)=f_j(\phi)$ the splitting between $c_i$ and
$c_j$ would be totally arbitrary and only $c_i+c_j$ could be
determined.

\noindent
{\bf Numerical analysis}

We apply the above procedure to the normalized lepton-energy
distributions derived in sec.2, eqs(\ref{single},\ref{double}). From
the theoretical point of view, perfectly-polarized initial beams
($P_{e^+}=P_{e^-}=\pm 1$) are the most attractive. However, those are
difficult to realize in practice, especially for the positron beam.
We shall therefore discuss the following two cases:

\hskip 2cm
(1)\ \ 
$P_{e^+}=0\,$\ \ vs\ \ $P_{e^-}=0,\ \pm 0.5,\ \pm 0.8$\ \ and\ \ $\pm
1$,

\hskip 2cm
(2)\ \ 
$P_{e^+}=P_{e^-}(\equiv P_e)=0,\ \pm 0.5,\ \pm 0.8$\ \ and\ \ $\pm
1$.

Before carrying out detailed computations, we shall briefly discuss
how the statistical errors ${\mit\Delta}c_i$ depend on $P_{e^\pm}$.
For this aim we have to check polarization effects in the lepton
spectra. These spectra depend on $P_{e^\pm}$ through the coefficients
$c_i$ and the functions $f_i$ in eqs.(\ref{fx},\ref{fxxbar}) as well,
but the strongest dependence comes from the normalization factor
since it is proportional to $\sigma_{e\bar{e}\to t\bar{t}}$ which is
\begin{eqnarray}
&&\sigma_{e\bar{e}\to t\bar{t}}
\sim (3-\beta^2)\Bigl[\:(1+P_{e^-}P_{e^+})D_V^{(0)}
-(P_{e^-}+P_{e^+})E_V^{(0)}\:\Bigr] \nonumber \\
&&\phantom{sigma_T}\ 
+2\beta^2 \Bigl[\:(1+P_{e^-}P_{e^+})D_A^{(0)}
-(P_{e^-}+P_{e^+})E_A^{(0)}\:\Bigr],
\end{eqnarray}
where $D_{V,A}^{(0)}$ and $E_{V,A}^{(0)}$ are the SM parts of
$D_{V,A}$ and $E_{V,A}$ in eq.(\ref{coefficient}). Neglecting the
vector-type part of $\gamma e\bar{e}$ coupling $v_e$ ($v_e=-1+4\sin^2
\theta_W$ is tiny for $\sin^2\theta_W=0.2315$), we have
$$
D_V^{(0)}=C(A_\gamma^2+A_Z^2 d'^2),\ \
D_A^{(0)}=CB_Z^2 d'^2,\ \
E_V^{(0)}=2CA_\gamma A_Z d',\ \
E_A^{(0)}=0.
$$
Since $E_V^{(0)}>0$ for $\sin^2\theta_W=0.2315$, negative
polarizations increase $\sigma_{e\bar{e}\to t\bar{t}}$. The matrix
$M_{ij}$ is proportional to $\sigma_{e\bar{e}\to t\bar{t}}$ through
the normalization factor, which means that negative polarizations
would reduce statistical errors, eq.(\ref{delc_i}), since the matrix 
$X_{ij} \propto 1/\sigma_{e\bar{e}\to t\bar{t}}$. As it has been
mentioned, $M_{ij}$ depends, to a certain extent, on $P_{e^\pm}$ also
through $c_i$ and $\eta^{(*)}$ in the functions $f_i$, therefore even
for nearly the same number of detected events (the same
$\sigma_{e\bar{e}\to t\bar{t}}$) statistical errors may differ.
However, the general tendency is consistent with this na\"{\i}ve
expectation as will be observed later in tables presenting our
results.

\noindent
{\bf 1. Single-distribution analysis}

First, we shall consider the single distribution. Using
eq.(\ref{delc_i}) for $d\sigma^\pm/dx$ we can obtain ${\mit\Delta}
c_{2,3}^\pm$, the statistical errors for the determination of
$c_{2,3}^\pm$, as a function of the expected number of detected
single-lepton events $N_{\ell}$. For a given integrated luminosity
$L$ and lepton-tagging efficiency $\epsilon_{\ell}$ one has $N_{\ell}
=B_{\ell}\sigma_{e\bar{e}\to t \bar{t}} L_{\rm eff}^{\ell}$, where
$L_{\rm eff}^{\ell} \equiv\epsilon_{\ell}L$ (in $\fbarn^{-1}$ units)
is the effective luminosity. In the following we use $\epsilon_{\ell}
=0.6$ and $L=100$~fb$^{-1}$ as an example of realistic experimental
constraint,\footnote{Assuming $L=100$~fb$^{-1}$ is in fact quite
    conservative since the integrated luminosity as high as $500$
    fb$^{-1}$ is being recently discussed \cite{luminosity} as a
    realistic possibility in the context of the TESLA collider design
	for $\sqrt{s}=500\gev$ .}\ 
and estimate $\sigma_{e\bar{e} \to t\bar{t}}$ within the SM by using
$\alpha(s) (\simeq 1/126)$.

\vskip 0.7cm
\renewcommand{\arraystretch}{1.4}
\begin{table}[h]
\bce 
\begin{tabular}{|c||c|c|c|c|c|c|c|}
\hline
(1) $P_{e^-}$ &  0   & +0.5 & +0.8 & +1.0 &$-$0.5&$-$0.8&$-$1.0\\
\hline
${\mit\Delta}c_2^\pm $ 
              & 0.13 & 0.16 & 0.12 & 0.09 & 0.09 & 0.08 & 0.07 \\
\hline
${\mit\Delta}c_3^\pm $ 
              & 0.08 & 0.10 & 0.08 & 0.06 & 0.06 & 0.05 & 0.05 \\
\hline
$N_{\ell}$      & 7676 & 6259 & 5409 & 4843 & 9093 & 9943 & 10509\\
\hline\hline
(2) $P_{e^{\phantom{-}}}$
              &  0   & +0.5 & +0.8 & +1.0 &$-$0.5&$-$0.8&$-$1.0\\
\hline
${\mit\Delta}c_2^\pm $ 
              & 0.13 & 0.11 & 0.08 & 0.07 & 0.07 & 0.05 & 0.05 \\
\hline
${\mit\Delta}c_3^\pm $ 
              & 0.08 & 0.07 & 0.05 & 0.04 & 0.05 & 0.04 & 0.03 \\
\hline
$N_{\ell}$      & 7676 & 6762 & 8055 & 9685 & 12429& 17122& 21019\\
\hline
\end{tabular}
\ece
\vspace*{-0.4cm}
\caption{Expected statistical errors in $c_{2,3}^\pm$ measurements
and the number of the single-lepton-inclusive events $N_{\ell}$ 
for beam polarization
(1) $P_{e^+}=0$ vs $P_{e^-}=0,\ \pm 0.5,\ \pm 0.8$ and $\pm 1$,
(2) $P_{e^+}=P_{e^-}(\equiv P_e)=0,\ \pm 0.5,\ \pm 0.8$ and $\pm 1$
at $\protect\sqrt{s}=500$~GeV. $N_{\ell}$ has been estimated within
the SM for $\epsilon_{\ell}=0.6$ and $L=100$ fb$^{-1}$.}    
\label{spol_errora}
\end{table} 

\vskip 0.7cm
In table~\ref{spol_errora} we present ${\mit\Delta}c_{2,3}^\pm$ and
$N_{\ell}$ for the above $\epsilon_{\ell}$ and $L$ with the described
configurations of beam polarization. From table \ref{spol_errora},
readers might conclude that the use of polarized beam(s) is quite
effective for providing higher precision, especially
negatively-polarized beams seem to be most suitable since we have
smaller ${\mit\Delta}c_{2,3}^\pm$ as anticipated in the above
discussion. Indeed, this is the case for $c_3^\pm$ measurement. For
instance, when ${\rm Re}(f_2^R),\:{\rm Re}(\bar{f}_2^L)=\pm 0.1$,
then $N_{S\!D}=|c_3^\pm|/{\mit\Delta}c_3^\pm$, statistical
significances for an observation of $c_3^\pm$, becomes 2.0 for
$P_{e^-}=-1$ and 3.3 for $P_e=-1$, which means we can expect $2
\sigma$ and $3\sigma$ confidence level respectively. However, for a
given set of non-standard couplings, the coefficients $c_2^\pm$ vary
with polarization. Therefore we should discuss their $N_{S\!D}$
inevitably instead of statistical errors only. We will consider the
following two sets of the couplings (of the order of $15\%$ of the
SM strength) in tables \ref{prea} and \ref{preb}:\footnote{One may
    notice that certain entries (some of $c_i$ coefficients) in
    tables \ref{prea} and \ref{preb} are identical. Indeed two
    polarization scenarios considered here provide for these cases
    exactly same values for $c_i$. Therefore, comparing statistical
    significances for them one can see the net effect of different
    statistics, as the expected number of events is different for the
    cases. The same will also apply to tables \ref{dprea} and
    \ref{dpreb}.} 
\begin{itemize}
\item[(a)] 
 ${\rm Re}(\delta\!A_{\gamma,Z})={\rm Re}(\delta\!B_{\gamma,Z})
 ={\rm Re}(\delta  C_{\gamma,Z})={\rm Re}(\delta\!D_{\gamma,Z})=0.1$,
\item[(b)] 
 ${\rm Re}(\delta\!A_\gamma)={\rm Re}(\delta\!B_\gamma)
 ={\rm Re}(\delta  C_\gamma)={\rm Re}(\delta\!D_\gamma)=0.1$,\\
 ${\rm Re}(\delta\!A_Z)={\rm Re}(\delta\!B_Z)
 ={\rm Re}(\delta  C_Z)={\rm Re}(\delta\!D_Z)=-0.1$.
\end{itemize}
\begin{table}[h]
\bce 
\begin{tabular}{|l||c|c|c|c|c|c|c|}
\hline
(1) $P_{e^-}$ &  0   & +0.5 & +0.8 & +1.0 &$-$0.5&$-$0.8&$-$1.0\\
\hline
\ \ \ $c_2^+$ & 0.39 & 0.36 & 0.28 & 0.17 & 0.38 & 0.36 & 0.34 \\
\hline
\ \ \ $c_2^-$ & 0.14 & 0.16 & 0.12 & 0.05 & 0.09 & 0.06 & 0.03 \\
\hline
$|c_2^+|/{\mit\Delta}c_2^\pm$
              & 3.03 & 2.31 & 2.25 & 1.83 & 4.10 & 4.65 & 4.96 \\
\hline
$|c_2^-|/{\mit\Delta}c_2^\pm$
              & 1.11 & 1.04 & 1.01 & 0.58 & 1.00 & 0.75 & 0.49 \\
\hline\hline
(2) $P_e$     &  0   & +0.5 & +0.8 & +1.0 &$-$0.5&$-$0.8&$-$1.0\\
\hline
\ \ \ $c_2^+$ & 0.39 & 0.28 & 0.19 & 0.17 & 0.36 & 0.35 & 0.34 \\
\hline
\ \ \ $c_2^-$ & 0.14 & 0.12 & 0.06 & 0.05 & 0.06 & 0.04 & 0.03 \\
\hline
$|c_2^+|/{\mit\Delta}c_2^\pm$
              & 3.03 & 2.52 & 2.47 & 2.59 & 5.20 & 6.30 & 7.01 \\
\hline
$|c_2^-|/{\mit\Delta}c_2^\pm$
              & 1.11 & 1.13 & 0.86 & 0.81 & 0.83 & 0.68 & 0.70 \\
\hline
\end{tabular}
\ece
\vspace*{-0.4cm}
\caption{Statistical significance of $c_2^\pm$ measurement for beam
polarization
(1) $P_{e^+}=0$ vs $P_{e^-}=0,\ \pm 0.5,\ \pm 0.8$ and $\pm 1$, and
(2) $P_{e^+}=P_{e^-}(\equiv P_e)=0,\ \pm 0.5,\ \pm 0.8$ and $\pm 1$,
and the parameter set (a)
${\rm Re}(\delta\!A_\gamma)={\rm Re}(\delta\!A_Z)
={\rm Re}(\delta\!B_\gamma)={\rm Re}(\delta\!B_Z)
={\rm Re}(\delta C_\gamma)={\rm Re}(\delta C_Z)
={\rm Re}(\delta\!D_\gamma)={\rm Re}(\delta\!D_Z)=0.1$
at $\protect\sqrt{s}=500$ GeV.}    
\label{prea}
\end{table}
\newpage
\begin{table}[h]
\bce 
\begin{tabular}{|l||c|c|c|c|c|c|c|}
\hline
(1) $P_{e^-}$ &  0   & +0.5 & +0.8 & +1.0 &$-$0.5&$-$0.8&$-$1.0\\
\hline
\ \ \ $c_2^+$ & 0.17 & 0.31 & 0.46 & 0.61 & 0.11 & 0.08 & 0.07 \\
\hline
\ \ \ $c_2^-$ &$-4\!\cdot\!10^{-3}$
                     & 0.04 & 0.11 & 0.19 &$-$0.01&$10^{-3}$
                                                        & 0.01 \\
\hline
$|c_2^+|/{\mit\Delta}c_2^\pm$
              & 1.33 & 1.97 & 3.70 & 6.63 & 1.15 & 1.07 & 1.02 \\
\hline
$|c_2^-|/{\mit\Delta}c_2^\pm$
              & 0.03 & 0.24 & 0.86 & 2.09 & 0.06 & 0.02 & 0.10 \\
\hline\hline
(2) $P_e$     &  0   & +0.5 & +0.8 & +1.0 &$-$0.5&$-$0.8&$-$1.0\\
\hline
\ \ \ $c_2^+$ & 0.17 & 0.46 & 0.59 & 0.61 & 0.08 & 0.07 & 0.07 \\
\hline
\ \ \ $c_2^-$ &$-4\!\cdot\!10^{-3}$
                     & 0.11 & 0.18 & 0.19 &$10^{-3}$
                                                 & 0.01 & 0.01 \\
\hline
$|c_2^+|/{\mit\Delta}c_2^\pm$
              & 1.33 & 4.14 & 7.86 & 9.38 & 1.20 & 1.31 & 1.44 \\
\hline
$|c_2^-|/{\mit\Delta}c_2^\pm$
              & 0.03 & 0.97 & 2.40 & 2.95 & 0.02 & 0.12 & 0.14 \\
\hline
\end{tabular}
\ece
\vspace*{-0.4cm}
\caption{Statistical significance of $c_2^\pm$ measurement for beam
polarization
(1) $P_{e^+}=0$ vs $P_{e^-}=0,\ \pm 0.5,\ \pm 0.8$ and $\pm 1$, and
(2) $P_{e^+}=P_{e^-}(\equiv P_e)=0,\ \pm 0.5,\ \pm 0.8$ and $\pm 1$,
and the parameter set (b)
${\rm Re}(\delta\!A_\gamma)=-{\rm Re}(\delta\!A_Z)
={\rm Re}(\delta\!B_\gamma)=-{\rm Re}(\delta\!B_Z)
={\rm Re}(\delta C_\gamma)=-{\rm Re}(\delta C_Z)
={\rm Re}(\delta\!D_\gamma)=-{\rm Re}(\delta\!D_Z)=0.1$
at $\protect\sqrt{s}=500$ GeV.}    
\label{preb}
\end{table}

These tables show that the use of negatively-polarized beam(s) is not
always optimal: for the parameter set (a) a good precision in $c_2^+$
measurement is realized when $P_e<0$, but even in this case the
precision in $c_2^-$ measurement becomes better for $P_e>0$ or even
$P_e=0$ (table \ref{prea}). Moreover in case (b) both $c_2^+$ and
$c_2^-$ get the highest precision for $P_e=+1$ (table \ref{preb}).
{\it Therefore one should carefully adjust optimal polarization to
test any given model of physics beyond the SM.} One can conclude (as
far as the coefficient sets discussed here are concerned) that the
appropriate beam polarization for the set (a) provides measurements
of $c_2^+$ at $5.0\sigma$ and $7.0\sigma$ level for $P_{e^-}=-1.0$
and $P_e=-1.0$, respectively. For the set (b) maximal statistical
significance for $c_2^+$ determination is $6.6$ and $9.4$ for
$P_{e^-}=+1.0$ and $P_e=+1.0$, respectively. Since $c_2^- \ll c_2^+$
it is seen that the maximal statistical significance for $c_2^-$ is
much lower: $1.1$ for the set (a) and $3.0$ for the set (b). 

\noindent
{\bf 2. Double-distribution analysis}

We can perform similar computations for the double-lepton
distribution. Results are presented in tables \ref{dpol_errora},
\ref{dprea} and \ref{dpreb}. We find again in table \ref{dpol_errora}
that negative polarizations give smaller ${\mit\Delta}c_i$. As a
result, $|c_{3,6}|/{\mit\Delta}c_{3,6}$ can be easily estimated from
this table once ${\rm Re}(f_2^R)$ and ${\rm Re}(\bar{f}_2^L)$ are
fixed. On the other hand, $c_{2,4,5}$ have polarization dependence
themselves, so we need tables \ref{dprea} and \ref{dpreb} in order to
draw a meaningful conclusion, where the statistical significance for
$c_{2,4,5}$ has been presented. Again some of $c_i$ in tables
\ref{dprea} and \ref{dpreb} are identical as in the case of the
single lepton channel.

\begin{table}[h]
\bce 
\begin{tabular}{|c||c|c|c|c|c|c|c|}
\hline
(1) $P_{e^-}$ &  0   & +0.5 & +0.8 & +1.0 &$-$0.5&$-$0.8&$-$1.0\\
\hline
${\mit\Delta}c_2 $ 
              & 0.20 & 0.23 & 0.21 & 0.17 & 0.16 & 0.14 & 0.13 \\
\hline
${\mit\Delta}c_3 $ 
              & 0.13 & 0.15 & 0.14 & 0.11 & 0.11 & 0.09 & 0.09 \\
\hline
${\mit\Delta}c_4 $ 
              & 0.31 & 0.35 & 0.39 & 0.41 & 0.30 & 0.29 & 0.28 \\
\hline
${\mit\Delta}c_5 $ 
              & 0.22 & 0.25 & 0.22 & 0.17 & 0.17 & 0.14 & 0.13 \\
\hline
${\mit\Delta}c_6 $ 
              & 0.14 & 0.16 & 0.14 & 0.12 & 0.11 & 0.09 & 0.09 \\
\hline
$N_{{\ell}{\ell}}$
              & 1013 &  826 &  714 &  639 & 1200 & 1312 & 1387 \\
\hline\hline
(2) $P_{e^{\phantom{-}}}$
              &  0   & +0.5 & +0.8 & +1.0 &$-$0.5&$-$0.8&$-$1.0\\
\hline
${\mit\Delta}c_2 $ 
              & 0.20 & 0.19 & 0.14 & 0.12 & 0.12 & 0.10 & 0.09 \\
\hline
${\mit\Delta}c_3 $ 
              & 0.13 & 0.12 & 0.09 & 0.08 & 0.08 & 0.07 & 0.06 \\
\hline
${\mit\Delta}c_4 $ 
              & 0.31 & 0.34 & 0.32 & 0.29 & 0.26 & 0.22 & 0.20 \\
\hline
${\mit\Delta}c_5 $ 
              & 0.22 & 0.19 & 0.14 & 0.12 & 0.13 & 0.10 & 0.09 \\
\hline
${\mit\Delta}c_6 $ 
              & 0.14 & 0.12 & 0.09 & 0.08 & 0.08 & 0.07 & 0.06 \\
\hline
$N_{{\ell}{\ell}}$
              & 1013 &  893 & 1063 & 1278 & 1641 & 2260 & 2775 \\
\hline
\end{tabular}
\ece
\vspace*{-0.4cm}
\caption{Expected statistical errors in $c_{2,3,4,5,6}$ measurements
and the expected observed numbers of the double-lepton-inclusive
events $N_{{\ell}{\ell}}$ for beam polarization
(1) $P_{e^+}=0$ vs $P_{e^-}=0,\ \pm 0.5,\ \pm 0.8$ and $\pm 1$,
(2) $P_{e^+}=P_{e^-}(\equiv P_e)=0,\ \pm 0.5,\ \pm 0.8$ and $\pm 1$
at $\protect\sqrt{s}=500$ GeV. $N_{{\ell}{\ell}}$ has been estimated
within the SM for $\epsilon_{\ell}=0.6$ and $L=100$ fb$^{-1}$.}    
\label{dpol_errora}
\end{table}

Among the coefficients for the double-leptonic spectrum, $c_{2,3}$
are $C\!P$-violating parameters. Since $c_3$ does not depend on the
beam polarization as already mentioned, one can just say (from table
\ref{dpol_errora}) that $3\sigma$ effects could be observed for $P_e=
-1.0$ if $\re(f_2^R-\bar{f}_2^L)/2>0.18$. On $c_2$ one has to
conclude from tables \ref{dprea} and \ref{dpreb} that for both sets
of non-standard couplings its determination would not be easy for the
assumed luminosity, as its statistical significance reaches at most
$1.7$. This is due to the smaller number of detected events in this
channel as it could have been anticipated. Still we can say that the
use of polarized beams is very helpful to increase precision. Indeed,
if we are able to achieve $L=500$~fb$^{-1}$ as discussed in
\cite{luminosity}, then $|c_2|/{\mit\Delta}c_2$ would reach $3.8$ for
$P_e=-1$ in case (a) (the same value could be obtained for $P_e=+1$
in case (b)), while we have only $|c_2|/{\mit\Delta}c_2=1.4$ if the
beams were unpolarized.

\begin{table}[h]
\bce 
\begin{tabular}{|l||c|c|c|c|c|c|c|}
\hline
(1) $P_{e^-}$
    &  0    & +0.5  & +0.8  & +1.0  &$-$0.5 &$-$0.8 &$-$1.0 \\
\hline
\ \ \ $c_2$
    &$-0.12$&$-$0.10&$-$0.08&$-$0.06&$-$0.14&$-$0.15&$-$0.16\\
\hline
\ \ \ $c_4$
    & 0.21  & 0.15  & 0.10  & 0.06  & 0.25  & 0.27  & 0.28  \\
\hline
\ \ \ $c_5$
    & 0.27  & 0.26  & 0.20  & 0.11  & 0.23  & 0.21  & 0.19  \\
\hline
$|c_2|/{\mit\Delta}c_2$
    & 0.61  & 0.42  & 0.37  & 0.34  & 0.89  & 1.08  & 1.21  \\
\hline
$|c_4|/{\mit\Delta}c_4$
    & 0.67  & 0.43  & 0.25  & 0.14  & 0.84  & 0.93  & 0.99  \\
\hline
$|c_5|/{\mit\Delta}c_5$
    & 1.24  & 1.05  & 0.92  & 0.63  & 1.41  & 1.45  & 1.44  \\
\hline\hline
(2) $P_e$
    & 0     & +0.5  & +0.8  & +1.0  &$-$0.5 &$-$0.8 &$-$1.0 \\
\hline
\ \ \ $c_2$
    &$-0.12$&$-$0.08&$-$0.06&$-$0.06&$-$0.15&$-$0.15&$-$0.16\\
\hline
\ \ \ $c_4$
    & 0.21  & 0.10  & 0.06  & 0.06  & 0.27  & 0.28  & 0.28  \\
\hline
\ \ \ $c_5$
    & 0.27  & 0.20  & 0.13  & 0.11  & 0.21  & 0.19  & 0.19  \\
\hline
$|c_2|/{\mit\Delta}c_2$
    & 0.61  & 0.41  & 0.44  & 0.48  & 1.21  & 1.53  & 1.71  \\
\hline
$|c_4|/{\mit\Delta}c_4$
    & 0.67  & 0.28  & 0.19  & 0.19  & 1.04  & 1.26  & 1.40  \\
\hline
$|c_5|/{\mit\Delta}c_5$
    & 1.24  & 1.03  & 0.89  & 0.90  & 1.62  & 1.85  & 2.04  \\
\hline
\end{tabular}
\ece
\caption{Statistical significance of $c_{2,4,5}$ measurement for beam
polarization
(1) $P_{e^+}=0$ vs $P_{e^-}=\pm 0.5,\ \pm 0.8$ and $\pm 1$, and
(2) $P_{e^+}=P_{e^-}(\equiv P_e)=\pm 0.5,\ \pm 0.8$ and $\pm 1$,
and the parameter set (a)
${\rm Re}(\delta\!A_\gamma)={\rm Re}(\delta\!A_Z)
={\rm Re}(\delta\!B_\gamma)={\rm Re}(\delta\!B_Z)
={\rm Re}(\delta  C_\gamma)={\rm Re}(\delta  C_Z)
={\rm Re}(\delta\!D_\gamma)={\rm Re}(\delta\!D_Z)=0.1$
at $\protect\sqrt{s}=500$ GeV.}
\label{dprea}
\end{table}

$c_{4,5,6}$ are $C\!P$-conserving coefficients. Concerning $c_6$,
$3\sigma$-level measurement is possible for $P_e=-1.0$ when
$\re(f_2^R+\bar{f}_2^L)/2>0.18$. On $c_4$ we are also led to a
similar conclusion to $c_2$, but $c_5$ determination is different.
That is, the statistical significance for $c_5$ measurement can reach
$2.0$ for $P_e=-1$ (case (a)) and $3.3$ for $P_e=+1$ (case (b)). This
is quite in contrast with that for $c_4$, which is less than 2 as one
can see from tables \ref{dprea} and \ref{dpreb}. 

\begin{table}[h]
\bce 
\begin{tabular}{|l||c|c|c|c|c|c|c|}
\hline
(1) $P_{e^-}$ &  0   & +0.5 & +0.8 & +1.0 &$-$0.5&$-$0.8&$-$1.0\\
\hline
\ \ \ $c_2$
    &$-0.09$&$-$0.14&$-$0.18&$-$0.21&$-$0.06&$-$0.04&$-$0.03\\
\hline
\ \ \ $c_4$
    & 0.11  & 0.14  & 0.18  & 0.20  & 0.08  & 0.07  & 0.06  \\
\hline
\ \ \ $c_5$
    & 0.08  &0.17   & 0.28  & 0.40  & 0.05  & 0.04  & 0.04  \\
\hline
$|c_2|/{\mit\Delta}c_2$
       & 0.43 & 0.58 & 0.84 & 1.22 & 0.35 & 0.29 & 0.25  \\
\hline
$|c_4|/{\mit\Delta}c_4$
       & 0.34 & 0.42 & 0.45 & 0.50 & 0.26 & 0.23 & 0.20  \\
\hline
$|c_5|/{\mit\Delta}c_5$
       & 0.39 & 0.69 & 1.29 & 2.30 & 0.30 & 0.29 & 0.30 \\
\hline\hline
(2) $P_e$ &  0   & +0.5 & +0.8 & +1.0 &$-$0.5&$-$0.8&$-$1.0\\
\hline
\ \ \ $c_2$
    &$-0.09$&$-$0.18&$-$0.21&$-$0.21&$-$0.04&$-$0.03&$-$0.03\\
\hline
\ \ \ $c_4$
    & 0.11  & 0.18  & 0.20  & 0.20  & 0.07  & 0.06  & 0.06  \\
\hline
\ \ \ $c_5$
    & 0.08  & 0.28  & 0.38  & 0.40  & 0.04  & 0.04  & 0.04  \\
\hline
$|c_2|/{\mit\Delta}c_2$
       & 0.43 & 0.94 & 1.49 & 1.73 & 0.33 & 0.33 & 0.35 \\
\hline
$|c_4|/{\mit\Delta}c_4$
       & 0.34 & 0.51 & 0.63 & 0.70 & 0.25 & 0.26 & 0.29 \\
\hline
$|c_5|/{\mit\Delta}c_5$
       & 0.39 & 1.45 & 2.72 & 3.25 & 0.33 & 0.38 & 0.42 \\
\hline
\end{tabular}
\vspace*{0.3cm}
\ece
\vspace*{-0.5cm}
\caption{Statistical significance of $c_{2,4,5}$ measurement for beam
polarization
(1) $P_{e^+}=0$ vs $P_{e^-}=\pm 0.5,\ \pm 0.8$ and $\pm 1$, and
(2) $P_{e^+}=P_{e^-}(\equiv P_e)=\pm 0.5,\ \pm 0.8$ and $\pm 1$,
and the parameter set (b)
${\rm Re}(\delta\!A_\gamma)=-{\rm Re}(\delta\!A_Z)
={\rm Re}(\delta\!B_\gamma)=-{\rm Re}(\delta\!B_Z)
={\rm Re}(\delta  C_\gamma)=-{\rm Re}(\delta  C_Z)
={\rm Re}(\delta\!D_\gamma)=-{\rm Re}(\delta\!D_Z)=0.1$
at $\protect\sqrt{s}=500$ GeV.}    
\label{dpreb}
\end{table}

\sec{Summary and comments}

Next-generation linear colliders of $e^+ e^-$, NLC, will provide a
cleanest environment for studying top-quark interactions. There, we
shall be able to perform detailed tests of the top-quark couplings to
the vector bosons and either confirm the SM simple
generation-repetition pattern or discover some non-standard
interactions. In this paper, assuming the most general
($\cp$-violating {\it and} $\cp$-conserving) couplings for $\gamma
\ttbar$, $Z \ttbar$ and $W tb$, we have studied in a
model-independent way the single- and the double-leptonic spectra
for arbitrary longitudinal beam polarizations. Then, the
optimal-observable technique \cite{optimalization} has been adopted
to determine non-standard couplings through measurements of these
spectra.

The method applied here, the optimal observables, allows to
disentangle various non-standard contributions to the production
process ($\eett$) and to the decay ($t \to W b$). Using the
single-leptonic-energy spectrum for $\ell^\pm$ and assuming
non-standard couplings of the order of $15\%$ of the SM strength, we
have found that an appropriate selection of the initial-beam
polarization may provide observable effects for non-standard
corrections to {\it the production process}, $|c_2^+|/{\mit\Delta}
c_2^+$, even at $9.4\sigma$ level when both $e^-$ and $e^+$ beams are
polarized and at $6.6\sigma$ when only $e^-$ beam is polarized. On
the other hand, from the same spectrum measurement one can expect on
{\it non-standard contributions to the top-quark decay} the
statistical significance of the signal $N_{S\!D}=|c_3^\pm|/
{\mit\Delta}c_3^\pm$ of the order of 3.0 and 2.0 for both beams
polarized and only electron beam polarized, respectively.

The direct application of the optimal method for the single spectrum
does not allow for discrimination between $\cp$-violating and
$\cp$-conserving non-standard interactions since their effects mix in
coefficients of the spectrum, $c_i^\pm$. However, as it was discussed
in ref.~\cite{GH_npb} one can easily combine measurements of the
spectrum for $\lp$ and $\lm$ in order to construct purely
$\cp$-violating observables.

In contrast with the single spectrum, utilizing the method of optimal
observables directly for the double-leptonic-energy spectrum one can
separately determine and disentangle the $\cp$-violating coupling
from the production of $\ttbar$ pairs ($c_2$) and the one from the
top-quark decay ($c_3$). For the typical strength of the non-standard
couplings discussed here, the highest statistical significance for
$\cp$ violation in {\it the production and/or in the decay} was
estimated to be $1.7$ for both beams polarized, while we found that
the maximal signal from $\cp$-conserving interactions in {\it the
production process} ($|c_5|/{\mit\Delta}c_5$) could reach $3.3$ and
$2.3$ for both and only electron beam polarized, respectively. For
$\cp$-conserving interactions in the decay the expected effect is
lower, namely $1.6$ for the statistical significance for both
considered cases of maximal polarization.

It should be emphasized that we have used in this study very
conservative integrated luminosity, namely $L=100$~fb$^{-1}$. That
is, the luminosity considered now as realistic is by factor $5$
larger. Therefore one may expect that even though we have not
considered any background here and our analysis does not take into
account any detector details (to a large extent they are not
available yet), the results presented here should serve as a fair
estimation of real signals for beyond-the-SM physics.

To summary, we found ($i$) the use of longitudinal beams could be
very effective in order to increase precision of the determination of
non-SM couplings, however ($ii$) optimal polarization depends on the
model of new physics under consideration, therefore polarization of
the initial beams should be carefully adjusted for each tested model.
{\it For such optimal polarization the maximal non-standard signal 
should be observable in the single-leptonic spectrum on the effects
generated by contributions (both $\cp$-conserving and
$\cp$-violating) to the production mechanism of $\ttbar$ pairs. On
the other hand, the most challenging measurement would be the
determination of $\cp$-conserving contributions to the decay
process.} Since we have already carried out a similar analysis of
possible consequences emerging from effective four-Fermi interactions
$e\bar{e}\to t\bar{t}$ and $t(\bar{t})\to b{\ell}^+
\nu(\bar{b}{\ell}^-\bar{\nu})$ in \cite{GHS}, this paper completes a
full analysis of modifications for lepton-energy distributions by
non-standard interactions of the top quark in a model-independent way
for polarized $e^+e^-$ experiments.

The results presented here are the most precise ones which could be
obtained from the single or double energy distribution alone. It will
of course be possible to achieve a higher precision by combing our
results with other statistically-independent data. Among them, lepton
angular distributions are very promising. Indeed what one could
measure via the energy spectra are the real parts of the non-standard
form factors, while we would be able to determine their imaginary
parts by using, e.g., an up-down asymmetry to the top direction as
shown in \cite{CKP}. However, non-SM effects in the decay process
were not taken into account in that study. The lepton angular
distributions relative to the initial beam direction will also give
us valuable information. Some analysis focusing on the $C\!P$
violation in the production vertices has been made in \cite{PR}.
However, comprehensive analysis including non-standard effects both
in the production and in the decay process for all measurable
distributions of the $\ttbar$ decay products seems to be needed
\cite{GH_new}.

Finally, let us give a brief comment on the effects of radiative
corrections. All the non-standard couplings considered here may be
generated at the multi-loop level within the SM. In fact,
$\cp$-violating couplings $\delta\!D_{\gamma,Z}$ and ${\rm Re}(f_2^R-
\bar{f}_2^L)$ requires at least two loops of the SM, so they are
negligible. However, $\cp$-conserving couplings $\delta\!
A_{\gamma,Z}$, $\delta\!B_{\gamma,Z}$, $\delta C_{\gamma,Z}$ and
${\rm Re}(f_2^R+\bar{f}_2^L)$ could be generated already at the
one-loop level approximation of QCD. Therefore, in order to
disentangle non-SM interactions and the one-loop QCD effects it is
important to calculate and subtract the QCD contributions from the
lepton-energy spectrum, this is however beyond the scope of this
paper.

\vspace*{0.6cm}
\centerline{ACKNOWLEDGMENTS}

\vspace*{0.3cm}
This work is supported in part by the State Committee for Scientific
Research (Poland) under grant 2 P03B 014 14 and by Maria
Sk\l odowska-Curie Joint Fund II (Poland-USA) under grant
MEN/NSF-96-252.

\vspace*{0.6cm}

\noindent \hspace*{-0.72cm}
{\bf Appendix}

The angular distribution of polarized $t\bar{t}$ pair is given by
the following formula:
\begin{eqnarray}
&&\frac{d\sigma}{d{\mit\Omega}_t}(e^+e^-\to t(s_+)\bar{t}(s_-))
\non\\
&&=\frac{3\beta\alpha^2}{16 s^3}\:\Bigl[
\:\:D_V\:[\:\{ 4m_t^2s+(lq)^2 \}(1-s_{+}s_{-})+s^2(1+s_{+}s_{-})
\non\\
&&\ \ \ \lspace +2s(ls_{+}\;ls_{-}-Ps_{+}\;Ps_{-})
           +\:2\,lq(ls_{+}\;Ps_{-}-ls_{-}\;Ps_{+})\:]
\non\\
&&\ \ \ +\:D_A\:[\:(lq)^2(1+s_{+}s_{-})-(4m_t^2s-s^2)(1-s_{+}s_{-})
\non\\
&&\ \ \ \lspace -2(s-4m_t^2)(ls_{+}\;ls_{-}-Ps_{+}\;Ps_{-})
           -\:2\,lq(ls_{+}\;Ps_{-}-ls_{-}\;Ps_{+})\:]
\non\\
&&\ \ \ 
-4\:{\rm Re}(D_{V\!\!A})\:m_t\,[\:s(\psp-\psm)+lq(\lsp+\lsm)\:]\non\\
&&\ \ \ +2\:{\rm Im}(D_{V\!\!A})\:[\:lq\,\eps(\splus,\smin,q,l)
+\lsm\eps(\splus,P,q,l)+\lsp\eps(\smin,P,q,l)\:]
\non\\
&&\ \ \ +4\:E_V\:\m s(\lsp+\lsm)+4\:E_A\:\m\,lq(\psp-\psm)
\non\\
&&\ \ \ 
+4\:{\rm Re}(E_{V\!\!A})\:[\:2\mq(\lsp\;\psm-\lsm\;\psp)-lq\:s\:]
\non\\
&&\ \ \ +4\:{\rm Im}(E_{V\!\!A})\:\m[\:\eps(\splus,P,q,l)
+\eps(\smin,P,q,l)\:]
\non\\
&&\ \ \ -\:{\rm Re}(F_1)\:\mr[\:lq\;s(\lsp-\lsm)
-\{(lq)^2+4\mq s\}(\psp+\psm)\:]
\non\\
&&\ \ \ +2\:{\rm Im}(F_1)\:[\:s\,\eps(\splus,\smin,P,q)
+lq\,\eps(\splus,\smin,P,l)\:]
\non\\
&&\ \ \ +2\:{\rm Re}(F_2)\:s(\psp\;\lsm+\psm\;\lsp)
\non\\
&&\ \ \ -\:{\rm Im}(F_2)\:\frac{s}{m_t}
[\:\eps(\splus,P,q,l)-\eps(\smin,P,q,l)\:]
\non\\
&&\ \ \ -2\:{\rm Re}(F_3)\:lq(\psp\;\lsm+\psm\;\lsp)
\non\\
&&\ \ \ +\:{\rm Im}(F_3)\:\frac{lq}{m_t}
[\:\eps(\splus,P,q,l)-\eps(\smin,P,q,l)\:]
\non\\
&&\ \ \ -\:{\rm Re}(F_4)\:\frac{s}{m_t}
[\:lq\;(\psp+\psm)-(s-4\mq)(\lsp-\lsm)\:]
\non\\
&&\ \ \ -2\:{\rm Im}(F_4)\:
[\:\psp\eps(\smin,P,q,l)+\psm\eps(\splus,P,q,l)\:]
\non\\
&&\ \ \ +2\:{\rm Re}(G_1)\:
[\:\{ 4\mq s+(lq)^2-s^2 \}(1-\sss)-2s\,\psp\psm
\non\\
&&\ \ \ \lspace +lq(\lsp\;\psm-\lsm\;\psp)\:]
\non\\
&&\ \ \ -\:{\rm Im}(G_1)\:\frac{lq}{m_t}
[\:\eps(\splus,P,q,l)+\eps(\smin,P,q,l)\:]
\non\\
&&\ \ \ -\:{\rm Re}(G_2)\:\frac{s}{m_t}
[\:(s-4\mq)(\lsp+\lsm)-lq\;(\psp-\psm)\:]
\non\\
&&\ \ \ -2\:{\rm Im}(G_2)\:
[\:\psp\eps(\smin,P,q,l)-\psm\eps(\splus,P,q,l)\:]
\non\\
&&\ \ \ -\:{\rm Re}(G_3)\:\frac{lq}{m_t}
[\:lq\;(\psp-\psm)-(s-4\mq)(\lsp+\lsm)\:]
\non\\
&&\ \ \ -2\:{\rm Im}(G_3)\:lq\,\eps(\splus,\smin,q,l)
\non\\
&&\ \ \ +2\:{\rm Re}(G_4)\:
[\:(s-4\mq)(\psp\;\lsm-\psm\;\lsp)+2\,lq\;\psp\psm\:]
\non\\
&&\ \ \ +\:{\rm Im}(G_4)\:\mr (s-4\mq)[\:\eps(\splus,P,q,l)
+\eps(\smin,P,q,l)\:]\:\:\Bigr],
\label{distribution}
\end{eqnarray}
where $\beta(\equiv\sqrt{1-4\mts/s})$ is the velocity of $t$ in
$\epem$ c.m. frame,
$$
P\equiv p_e+p_{\bar{e}}\:(=p_t+p_{\bar{t}})\,,\ \ l\equiv
p_e-p_{\bar{e}}\,,\ \ q\equiv p_t-p_{\bar{t}}\,,
$$
the symbol $\epsilon(a,b,c,d)$ means $\epsilon_{\mu\nu\rho\sigma}
a^\mu b^\nu c^\rho d^\sigma$ for $\epsilon_{0123}=+1$,
\begin{eqnarray}
&& D_V\:\equiv \:C\,[\,\avs-2\av\az\ci+\azs\cz+2(\av-\az\ci)
{\rm Re}(\delta\!A_\gamma)
\non\\
&& \lspace-2\{ \av\ci-\az\cz \}{\rm Re}(\delta\!A_Z)\,],
\non\\
&& D_A\:\equiv \:C\,[\,B_Z^2(1+v_e^2)\cci^2
-2B_Z v_e \cci{\rm Re}(\delta\!B_\gamma)
+2B_Z(1+v_e^2)\cci^2{\rm Re}(\delta\!B_Z)\,],
\non\\
&& D_{V\!\!A}\:\equiv \:C\,
[\,-\av\bz\ci+\az\bz\cz-\bz\ci(\delta\!A_\gamma)^*
\non\\
&& \lspace
+(A_\gamma -v_e \cci A_Z)\delta\!B_\gamma +\bz\cz(\delta\!A_Z)^* 
\non\\
&& \lspace-\{ \av\ci-\az\cz \}\delta\!B_Z \,],
\non\\
&& E_V\:\equiv \:2C\,
[\,\av\az\cci-\azs\ccz+\az\cci{\rm Re}(\delta\!A_\gamma)
+(\av\cci-2\az\ccz){\rm Re}(\delta\!A_Z) \,],
\non\\
&& E_A\:\equiv \:2C\,[\,-\bzs\ccz+B_Z \cci{\rm Re}(\delta\!B_\gamma)
-2\bz\ccz{\rm Re}(\delta\!B_Z) \,],
\non\\
&& E_{V\!\!A}\:\equiv \:C\,
[\,\av\bz\cci-2\az\bz\ccz+\bz\cci(\delta\!A_\gamma)^*
+A_Z \cci\delta\!B_\gamma
\non\\
&& \lspace-2\bz\ccz(\delta\!A_Z)^*
+(\av\cci-2\az\ccz)\delta\!B_Z \,],
\non\\
&& F_1\:\equiv \:C\,[\, -(\av-\az\ci)\delta\!D_\gamma
     +\{\av\ci-\az\cz\}\delta\!D_Z \,],
\non\\
&& F_2\:\equiv \:C\,[\, -\az\cci\delta\!D_\gamma
     -(\av\cci-2\az\ccz)\delta\!D_Z \,],
\non\\
&& F_3\:\equiv \:C\,[\, \bz\ci\delta\!D_\gamma-\bz\cz\delta\!D_Z \,],
\non\\
&& F_4\:\equiv \:C\,
[\, -\bz\cci\delta\!D_\gamma+2\bz\ccz\delta\!D_Z \,],
\non\\
&& G_1\:\equiv \:C\,[\, (\av-\az\ci)\delta C_\gamma
     -\{\av\ci-\az\cz\}\delta C_Z \,],
\non\\
&& G_2\:\equiv \:C\,[\, \az\cci\delta C_\gamma
     +(\av\cci-2\az\ccz)\delta C_Z \,],
\non\\
&& G_3\:\equiv \:C\,[\, -\bz\ci\delta C_\gamma +\bz\cz\delta C_Z \,],
\non\\
&& G_4\:\equiv \:C\,
[\, \bz\cci\delta C_\gamma-2\bz\ccz\delta C_Z \,]
\label{coefficient}
\end{eqnarray}
and
$$
C \equiv 1/(4\sin^2\theta_W),\ \ \
\cci\equiv s/[4\sin\theta_W\cos\theta_W(s-M_Z^2)].
$$
In the above formulas, only linear terms in non-standard couplings
have been kept. 

The functions $f(x)$, $g(x)$, $\delta\! f(x)$ and $\delta g(x)$ in
eqs.(\ref{fx}) and (\ref{fxxbar}) are given as
\begin{eqnarray*}
&&f(x)=C_1 \Bigl\{\; r(r-2)+2x{{1+\beta}\over{1-\beta}}
-x^2 \Bigl({{1+\beta}\over{1-\beta}}\Bigr)^2\; \Bigr\}, \\
&&\hspace*{6.5cm}({\rm for\ the\ interval}\ I_1,\ I_4) \\
&&\phantom{f(x)}=C_1 \, (1-r)^2,
  \hspace*{3.02cm}({\rm for\ the\ interval}\ I_2) \\
&&\phantom{f(x)}=C_1 \, (1-x)^2,
  \hspace*{2.98cm}({\rm for\ the\ interval}\ I_3,\ I_6) \\
&&\phantom{f(x)}=C_1 \, x\Bigl\{\; x +{{4\beta}\over{1-\beta}}
-x \Bigl({{1+\beta}\over{1-\beta}}\Bigr)^2\; \Bigr\}, \\
&&\hspace*{6.5cm}({\rm for\ the\ interval}\ I_5)
\end{eqnarray*}

\begin{eqnarray*}
&&g(x)=C_2 \Bigl[\; -rx +x^2 {{1+\beta}\over{1-\beta}}
-x\ln {{x(1+\beta)}\over{r(1-\beta)}} \\
&&\phantom{g(x)}\ \ \ \ \ \ \ \
+{1\over{2(1+\beta)}}\Bigl\{\; r(r-2)+2x{{1+\beta}\over{1-\beta}}
-x^2 \Bigl({{1+\beta}\over{1-\beta}}\Bigr)^2\; \Bigr\}\; \Bigr], \\
&&\hspace*{6.5cm}({\rm for\ the\ interval}\ I_1,\ I_4) \\
&&\phantom{g(x)}
=C_2 \Bigl\{\; (1-r+\ln r)x +{1\over{2(1+\beta)}}(1-r)^2 \;\Bigr\},\\
&&\hspace*{6.5cm}({\rm for\ the\ interval}\ I_2) \\
&&\phantom{g(x)}
=C_2 \Bigl\{\; (1-x+\ln x)x +{1\over{2(1+\beta)}}(1-x)^2 \;\Bigr\},\\
&&\hspace*{6.5cm}({\rm for\ the\ interval}\ I_3,\ I_6) \\
&&\phantom{g(x)}
=C_2 \,x\Bigl[\; {{2\beta x}\over{1-\beta}}
-\ln{{1+\beta}\over{1-\beta}} \\
&&\phantom{g(x)}\ \ \ \ \ \ \ \
+{1\over{2(1+\beta)}}\Bigl\{\; x+{{4\beta}\over{1-\beta}}
-x \Bigl({{1+\beta}\over{1-\beta}}\Bigr)^2\; \Bigr\}\; \Bigr], \\
&&\hspace*{6.5cm}({\rm for\ the\ interval}\ I_5)
\end{eqnarray*}
where
$$
C_1\equiv {3\over{2W}}{{1+\beta}\over\beta},\ \ \ \ \ \ \ \
C_2\equiv {3\over{W}}{{(1+\beta)^2}\over\beta},
$$
and $W\equiv (1-r)^2(1+2r)$ with $r \equiv (M_W/m_t)^2$ as defined in
the main text,
\begin{eqnarray*}
&&\delta\! f(x)
=C_3 \Bigl\{\; {1\over 2}r(r+8)-2x(r+2){{1+\beta}\over{1-\beta}}
+{3\over 2}x^2 \Bigl({{1+\beta}\over{1-\beta}}\Bigr)^2 \\
&&\phantom{\delta\! f(x)}\ \ \ \ \ \ \ \
+(1+2r)\ln{{x(1+\beta)}\over{r(1-\beta)}}\; \Bigr\}, \\
&&\hspace*{6.5cm}({\rm for\ the\ interval}\ I_1,\ I_4) \\
&&\phantom{\delta\! f(x)}
=C_3 \Bigl\{\; {1\over 2}(r-1)(r+5)-(1+2r)\ln r\; \Bigr\}, \\
&&\hspace*{6.5cm}({\rm for\ the\ interval}\ I_2) \\
&&\phantom{\delta\! f(x)}
=C_3 \Bigl\{\; {1\over 2}(x-1)(5+4r-3x)-(1+2r)\ln x\; \Bigr\}, \\
&&\hspace*{6.5cm}({\rm for\ the\ interval}\ I_3,\ I_6) \\
&&\phantom{\delta\! f(x)}
=C_3 \Bigl\{\; (1+2r)\ln{{1+\beta}\over{1-\beta}}
-{{4\beta x}\over{1-\beta}}(r+2)
+{{6\beta}\over{(1-\beta)^2}}x^2 \; \Bigr\}, \\
&&\hspace*{6.5cm}({\rm for\ the\ interval}\ I_5)
\end{eqnarray*}
\begin{eqnarray*}
&&\delta g(x)
=C_3 \Bigl[\; 1-\beta +2(3-\beta)r+{1\over 2}r^2
-{3\over 2}(1-2\beta)\Bigl({{1+\beta}\over{1-\beta}}\Bigr)^2 x^2 \\
&&\phantom{\delta\! f(x)}\ \ \ \ \ \ \ \
+(1+\beta)x\Bigl\{ {1\over r}(r-1)(3r+1)-{{2(r+2)}\over{1-\beta}}
\Bigr\} \\
&&\phantom{\delta\! f(x)}\ \ \ \ \ \ \ \
+\{ 1+2r+2(1+\beta)(r+2)x \}
\ln{{x(1+\beta)}\over{r(1-\beta)}}\; \Bigr], \\
&&\hspace*{6.5cm}({\rm for\ the\ interval}\ I_1,\ I_4) \\
&&\phantom{\delta\! f(x)}
=C_3 \Bigl[\; {1\over 2}(r-1)(r+5)-(1+2r)\ln r \\
&&\phantom{\delta\! f(x)}\ \ \ \ \ \ \ \
+(1+\beta)x\Bigl\{ {1\over r}(r-1)(5r+1)-2(r+2)\ln r \Bigr\}\;
\Bigr], \\
&&\hspace*{6.5cm}({\rm for\ the\ interval}\ I_2) \\
&&\phantom{\delta\! f(x)}
=C_3 \Bigl[\; -{7\over 2}-4r-\beta (2r+1)+2x\{1-\beta+r(2+\beta)\} \\
&&\phantom{\delta\! f(x)}\ \ \ \ \ \ \ \
+{3\over 2}(1+2\beta)x^2-\{2r+1+2(1+\beta)(r+2)x \}\ln x\; \Bigr], \\
&&\hspace*{6.5cm}({\rm for\ the\ interval}\ I_3,\ I_6) \\
&&\phantom{\delta\! f(x)}
=C_3 \Bigl[\; -(1+2r)\Bigl(2\beta-\ln{{1+\beta}\over{1-\beta}}\Bigr)
+{{6\beta^3}\over{(1-\beta)^2}}x^2 \\
&&\phantom{\delta\! f(x)}\ \ \ \ \ \ \ \
-2(r+2)x\Bigl\{ {{2\beta}\over{1-\beta}}
-(1+\beta)\ln{{1+\beta}\over{1-\beta}}\Bigr\} \; \Bigr], \\
&&\hspace*{6.5cm}({\rm for\ the\ interval}\ I_5)
\end{eqnarray*}
where
$$
C_3 \equiv {6\over W}{{1+\beta}\over\beta}{\sqrt{r}\over{1+2r}}.
$$
The intervals $I_i$ ($i=1\sim 6$) of $x$ are given by
\begin{eqnarray*}
&&I_1:\ \ r(1-\beta)/(1+\beta) \leq x \leq (1-\beta)/(1+\beta), \\
&&I_2:\ \ (1-\beta)/(1+\beta)  \leq x \leq r,                   \\
&&I_3:\ \ r                    \leq x \leq 1, \\
&&\ \ \ \ \ \ \ \ \ \ \ \ \ \
        (I_{1,2,3}\ {\rm are \ for}\ r\geq (1-\beta)/(1+\beta)) \\
&&I_4:\ \ r(1-\beta)/(1+\beta) \leq x \leq r,                   \\
&&I_5:\ \ r                    \leq x \leq (1-\beta)/(1+\beta),\\
&&I_6:\ \ (1-\beta)/(1+\beta)  \leq x \leq 1.                   \\
&&\ \ \ \ \ \ \ \ \ \ \ \ \ \
        (I_{4,5,6}\ {\rm are \ for}\ r\leq (1-\beta)/(1+\beta))
\end{eqnarray*}
\newpage
%


\begin{thebibliography}{99}
%
\bibitem{Bigi}
I. Bigi and H. Krasemann, \zfp{7}{1981}{127};\\
J.H. K\"uhn, {\it Acta Phys. Austr. Suppl.} XXIV (1982), 203;\\
I. Bigi, Yu. Dokshitser, V. Khoze, J.H. K\"uhn and P. Zerwas,
\plb{181}{1986}{157}.
%
\bibitem{data97}
All electroweak data used here are taken from: \\
Talks by G. Altarelli, by P. Giromini, by Y.Y. Kim, and by
J. Timmermans at {\it XVIII International Symposium on Lepton-Photon
Interactions}, Jul.28 - Aug.1, 1997, Hamburg, Germany.
%
\bibitem{SP}
C.R. Schmidt and M.E. Peskin, \prl{69}{1992}{410}.
%
\bibitem{CKP}
D. Chang, W.-Y. Keung and I. Phillips, \npb{408}{1993}{286}
(hep-ph/9301259); {\it ibid}. {\bf B429}(1994), 255(Erratum).
%
\bibitem{AS}
T. Arens and L.M. Sehgal, \prd{50}{1994}{4372}.
%
\bibitem{BG}
\bg, \plb{305}{1993}{384}.
%
\bibitem{GH_npb}
\bg\ and Z. Hioki, \npb{484}{1997}{17} (hep-ph/9604301).
%
\bibitem{GH_plb}
\bg\ and Z. Hioki, \plb{391}{1997}{172} (hep-ph/9608306).
%
\bibitem{BGH}
L. Brzezi\'nski, \bg\ and Z. Hioki, \ijmpa{14}{1999}{1261}
(hep-ph/9710358). 
%
\bibitem{GHS}
\bg, Z. Hioki and M. Szafranski, \prd{58}{1998}{035002}
(hep-ph/9712357).
%
\bibitem{Rest}
C.A. Nelson, \prd{41}{1990}{2805};\\
W. Bernreuther and O. Nachtmann, \plb{268}{1991}{424};\\
W. Bernreuther, T. Schr\"{o}der and T.N. Pham,
\plb{279}{1992}{389};\\
R. Cruz, \bg\, and J.F. Gunion, \plb{289}{1992}{440};\\
D. Atwood and A. Soni, \prd{45}{1992}{2405};\\
G.L. Kane, G.A. Ladinsky, and C.-P. Yuan, \prd{45}{1992}{124};\\  
W. Bernreuther, J.P. Ma, and T. Schr\"{o}der, \plb{297}{1992}{318};\\
\bg\ and W.-Y. Keung, \plb{316}{1993}{137};\\
D. Chang, W.-Y. Keung, and I. Phillips, \prd{48}{1993}{3225};\\
G.A. Ladinsky and C.-P. Yuan, \prd{49}{1994}{4415};\\
W. Bernreuther and P. Overmann, \zfp{61}{1994}{599}; {\it ibid}.
{\bf C72} (1996), 461 (hep-ph/9511256);\\
F. Cuypers and S.D. Rindani, \plb{343}{1995}{333} (hep-ph/9409243);\\
P. Poulose and S.D. Rindani, \plb{349}{1995}{379} (hep-ph/9410357);
\prd{54}{1996}{4326} (hep-ph/9509299); \plb{383}{1996}{212}
(hep-ph/9606356);\\
J.M. Yang and B.-L. Young, \prd{56}{1997}{5907} (hep-ph/9703463); \\
M.S. Baek, S.Y. Choi and C.S. Kim, \prd{56}{1997}{6835} 
(hep-ph/9704312); \\ 
A. Bartl, E. Christova, T. Gajdosik and W. Majerotto,
Report HEPHY-PUB-684 (hep-ph/9802352); Report HEPHY-PUB-686
(hep-ph/9803426).
%
\bibitem{effective_lag} W. Buchm\"uller and D. Wyler,
\npb{268}{1986}{621}.\\
See also \\
C.J.C. Burges and H.J. Schnitzer, \npb{228}{1983}{464}; \\
C.N. Leung, S.T. Love and S. Rao, \zfp{31}{1986}{433};\\
C. Arzt, M.B. Einhorn and J. Wudka, \npb{433}{1995}{41}
(hep-ph/9405214).
%
\bibitem{optimalization} 
J.F. Gunion, \bg\ and X-G. He, \prl{77}{1996}{5172}
(hep/ph-9605326). \\
See also \\
D. Atwood and A. Soni, in \cite{Rest};\\
M. Davier, L. Duflot, F. Le Diberder and A. Roug\'{e},
\plb{306}{1993}{411};\\
M. Diehl and O. Nachtmann, \zfp{62}{1994}{397};\\
\bg\ and J.F. Gunion, \plb{350}{1995}{218} (hep-ph/9501339).
%
\bibitem{technique}
Y.S. Tsai, \prd{4}{1971}{2821};\\  
S. Kawasaki, T. Shirafuji and S.Y. Tsai, \ptp{49}{1973}{1656}.
%
\bibitem{HO}
Z. Hioki and K. Ohkuma, \prd{59}{1999}{037503}.
%
\bibitem{Tsai}
Y.S. Tsai, \prd{51}{1995}{3172}.
%
\bibitem{Blon}
A. Blondel, \plb{202}{1988}{145}.
%
\bibitem{cprelation} 
W. Bernreuther, O. Nachtmann, P. Overmann and T. Schr\"{o}der,
\npb{388}{1992}{53};\\
\bg\ and J.F. Gunion, \plb{287}{1992}{237}.
%
\bibitem{luminosity}
R. Brinkmann, Talk at ECFA/DESY Linear Collider Workshop {\it
2nd ECFA/DESY Study on Physics and Detectors for a Linear 
Electron-Positron Collider}, LAL, Orsay, France, April 5-7, 1998.
%
\bibitem{PR}
P. Poulose and S.D. Rindani in \cite{Rest}.
%
\bibitem{GH_new}
\bg\ and Z. Hioki, work in progress.
%
\end{thebibliography}
\end{document}